\documentclass[12pt]{iopart}
\usepackage[pdftex]{graphicx}
\usepackage{placeins}
\usepackage{cite}
\begin{document}
\title[Coarse-graining of the dynamics of entangled polymer melts]
   {\review[Coarse-graining of the dynamics of entangled polymer melts]{Systematic coarse-graining of the dynamics of entangled polymer melts: the road from chemistry to rheology}}

\author{JT Padding$^{1,2}$\footnote[1]{e-mail adress: \tt <j.t.padding@gmail.com>} and WJ Briels$^2$\footnote[2]{e-mail adress: \tt <w.j.briels@utwente.nl>}}
\address{$^{1}$ Institut de la Mati\`ere Condens\'ee et des Nanosciences, Universit\'e catholique de Louvain, Croix du Sud 1, 1348 Louvain-la-Neuve, Belgium}
\address{$^2$ Computational Biophysics, Univ.~of Twente, PO Box 217, 7500 AE Enschede, The Netherlands}

\date{\today}
\begin{abstract} 	

For optimal processing and design of entangled polymeric materials
it is important to establish a rigorous link between the detailed molecular composition of the polymer
and the viscoelastic properties of the macroscopic melt. 
We review current and past computer simulation techniques and critically
assess their ability to provide such a link between chemistry and rheology.
We distinguish between two classes of coarse-graining levels, which we term coarse-grained molecular dynamics
(CGMD) and coarse-grained stochastic dynamics (CGSD). In CGMD the coarse-grained beads are
still relatively hard, thus automatically preventing bond crossing. This also implies an upper limit on the
number of atoms that can be lumped together (up to five backbone carbon atoms) 
and therefore on the longest chain lengths that can be studied.
To reach a higher degree of coarse-graining, in CGSD many more atoms are lumped together (more than
ten backbone carbon atoms), leading to
relatively soft beads. In that case friction and stochastic forces dominate the interactions, and actions must
be undertaken to prevent bond crossing. We also review alternative methods that make use of the tube model of polymer dynamics,
by obtaining the entanglement characteristics through a primitive path analysis and by simulation of a primitive
chain network. We finally review super-coarse-grained methods in which an entire polymer is
represented by a single particle, and comment on ways to include memory effects and transient forces.

\end{abstract}

\pacs{66.20.Cy,47.57.Ng,83.80.Sg}




\section{Introduction}

Ever since the first synthetic polymers were made by Staudinger and their molecular weights
were measured by Debye and Bueche, polymer systems have received continual interest from
both theorists and experimentalists. 
Most processing of polymers takes place
in the melt state, when they are very viscous
and have surprising (temporary) elastic properties.
For optimal processing, design, and application of polymeric melts it is
important to establish a rigorous link between their molecular
composition and macroscopic mechanical properties, i.e. between polymer
chemistry and rheology.

Establishing such a link is by no means an easy task and has despite considerable progress
not yet been fully achieved. Unlike simple molecules, interactions
within and between polymer chains are characterised by a large range of different time and
length scales. The dynamics of atomic bond vibrations are characterised
by {\AA}ngstrom and sub-picosecond scales, whereas the dynamics of statistically
independent Kuhn segments are characterised by nanometer and tens of picosecond scales \cite{DoiEdwards}.
Beyond this, the connectivity and mutual uncrossability of the segments cause an
interdependence between features on a hierarchy of scales.
The full chain, finally, is characterised by its radius of gyration of
10 to 100 nanometers. This may not appear as a very large increase from the Kuhn segment scale, but 
the associated time scales increase dramatically, to milliseconds, seconds or even longer.

The enormously long intrinsic time scales are often rationalized by viewing polymer systems as temporary rubbery networks.
Such a network arises as a result of mutual uncrossability of the polymer chains - they are \textit{entangled}.
With the advent of reptation theory of de Gennes, Doi and Edwards \cite{DoiEdwards}, a new concept
was introduced in the theory of polymer dynamics. In reptation theory
each polymer is supposed to move in a tube around a Gaussian path in space.
The tube only serves one purpose, namely to roughly represent the uncrossability
of the surrounding chains and to turn a difficult multichain problem into
a one-chain problem. The tube clearly is an idealised mean field concept, and
many extensions of the original model have been postulated to explain experimental
observations; the difficulty is that without detailed information such postulates 
often remain unchecked \cite{Likhtman2009}.

\begin{figure}[tp]
\centering
  \scalebox{0.7}{\includegraphics{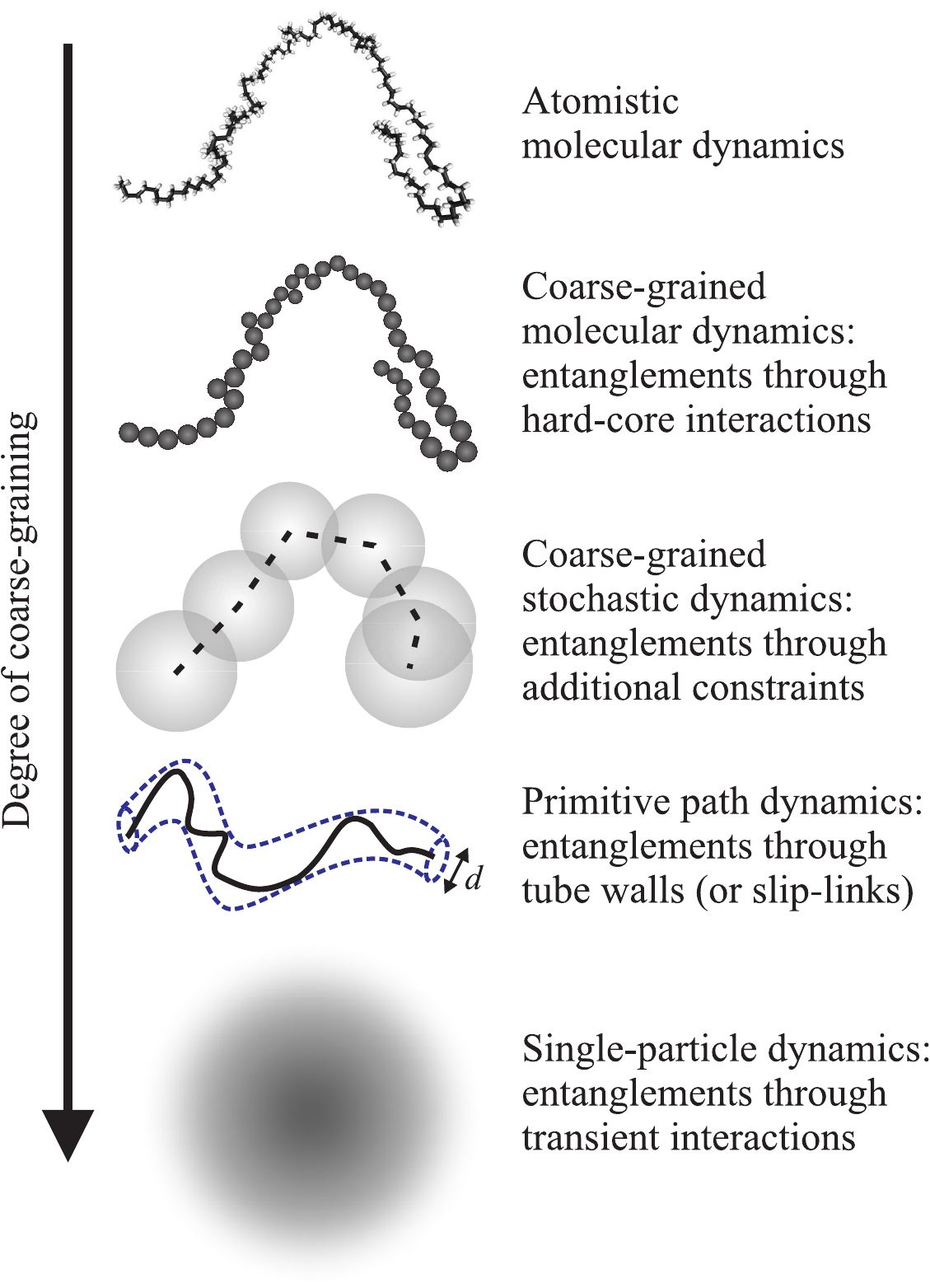}}
  \caption{The dynamics of polymers can be simulated by different methods which differ in their level of detail
  describing polymer configurations and entanglement effects. 
  \textit{Top}: atomistic molecular dynamics simulations are usually sufficiently detailed to faithfully predict the dynamic
  properties of polymers of a specified chemistry. \textit{Second from top}: small groups of atoms are lumped together (up to five backbone carbon atoms) into coarse-grained entities;
  the interaction between these `beads' are still hard enough to make bond crossing
  energetically unfavourable. \textit{Middle}: many more atoms are lumped together (more than 10 backbone
  carbon atoms) into coarse-grained entities; the interactions
  between these `blobs' are so soft that uncrossability needs to be enforced by additional constraints.
  \textit{Second from bottom}: polymer chains are forced to diffuse along primitive paths, representing the temporary
  topological network which arises because of uncrossability of chains. \textit{Bottom}: each polymer chain is represented by
  just one particle; the entanglement effect is captured through addition of slow variables whose deviations from equilibrium lead to transient forces.
  \label{fig_CGlevels}}
\end{figure}

In this review we focus on computer simulation techniques that may help establishing the link
between chemistry and large scale dynamics and rheology. Computer simulations
can check the assumptions made in theoretical models, and either accept or reject them \cite{Likhtman2009}.
Ideally, one would like to use atomistically detailed models because they can accurately capture the subtleties in the 
interactions between the polymers. However, because of the intrinsically slow dynamics, a direct
prediction of the large scale dynamics and rheology from such detailed models is computationally unattainable
(except for the case of unentangled or slightly entangled polymers). Traversing the road from chemistry to rheology is
usually a bumpy ride: the only way to reach the larger scales is by going through a succession of coarse-graining steps.
Different levels of coarse-graining are schematically indicated in Fig.~\ref{fig_CGlevels}.

A number of excellent reviews on coarse-graining of entangled polymers have already appeared in the literature. Some of these
reviews focus mainly on static properties \cite{Baschnagel2000,Kremer2002,Faller2004,Aleman2009}, whereas others
focus on the dynamic properties as well \cite{Muller-Plathe2002,Muller-Plathe2003,Paul2004}.
This review provides an up-to-date overview of past and current coarse-graining efforts, with a
strong emphasis on our own opinion about the correct way to proceed in the future. We therefore
focus on methods to handle friction, uncrossability and memory, all of which are intimately 
connected to the long time dynamic properties and rheology.

The review is structured as follows. In section \ref{sec_atomistic} we review the work that has been done using
atomistically detailed models. In section \ref{sec_theory} we introduce the theory of coarse-graining and make
a classification of coarse-grained dynamics methods into two classes: 1) those with relatively low levels of coarse-graining (up to five backbone carbon atoms),
where friction is not dominant or ignored and where uncrossability is included automatically by the excluded volume
interactions between relatively hard beads, and 2) those with relatively high levels of coarse-graining (more than ten backbone carbon atoms),
where friction is dominant and measures need to be taken to ensure uncrossability.
Section \ref{sec_cgmd} reviews the work that has been undertaken in the first class and section \ref{sec_cgsd} reviews the
work done in the second class. In section \ref{sec_primitive} we review the attempts that have been undertaken at characterising
primitive paths and the use of primitive primitive path models.
In section \ref{sec_super} we focus on the ultimate form of particulate coarse-graining,
where an entire polymer is represented by just one particle. Finally, we conclude in section \ref{sec_concl}.

%

\section{Atomistically detailed models}
\label{sec_atomistic}

\subsection{Possibilities and limitations}

On a microscopic level, molecular dynamics (MD) and Monte Carlo (MC) simulations can be performed,
in which each atom of a polymer chain is represented separately, see
Fig.~\ref{fig_CGlevels} (top) \cite{Padding2007}. The atoms are modelled as interacting particles and their positions
are updated according to Newton's laws (MD) or by certain trial moves which are accepted, depending on the statistical
ensemble, with a certain probability (MC).
Accurate force fields have been constructed to cater for anyone's research interests,
provided they exclude chemical reactions and other phenomena of a quantum mechanical nature.
Bulk behaviour is simulated by applying periodic boundary conditions to the simulation box.
Typical MD simulations cover the motion of a few tens of thousands of atoms over a period of a few nanoseconds;
on current computers such a run would take a week to complete.

This already sets a limit to the length of polymers that can be simulated in atomistic detail.
First, the radius of gyration of the polymer chain typically grows as $\sqrt{n}$,
where $n$ is the number of monomers per chain. A polymer should not interact
with itself via the periodic boundaries, which means that the volume of the box, and hence
the number of particles, should scale as $n^{3/2}$. Second, the longest intrinsic relaxation time of
a polymer chain in a melt scales very fast, usually as $n^2$ for unentangled polymers and as
$n^{3.4}$ for entangled polymers \cite{DoiEdwards}. To measure long-time correlation functions,
the simulations must be performed at least as long as this time scale.

Sometimes one is interested in the static properties of a melt of long polymer chains,
or in faster dynamical relaxation processes on a more local scale. In such cases, one still needs
to make sure that the atomistic conformations in the simulation box have sufficiently relaxed
to faithfully represent an equilibrium state.
 	 	
\subsection{Equilibration schemes for atomistically detailed models}

So how does one ensure that an atomistically detailed polymer melt is sufficiently relaxed? A first reaction may be
to start with an ensemble of chains with a correct end-to-end distance distribution, arrange them randomly in the
simulation box, and introduce excluded volume rapidly. However, Auhl et al. \cite{Auhl2003} showed that this procedure leads
to deformations on short length scales, which relax only when the chains have moved over their own size, i.e. after
one longest intrinsic relaxation time. The authors also showed how this local deformation may be overcome by first
pre-packing the Gaussian chains, reducing the density fluctuations in the system, followed by a more gradual introduction
of excluded volume. Another method is to apply a double-bridging Monte Carlo algorithm in which new bonds are formed
across a pair of chains, creating two new chains substantially different from the original two \cite{Auhl2003}.

\begin{figure}[tp]
\centering
  \scalebox{0.6}{\includegraphics{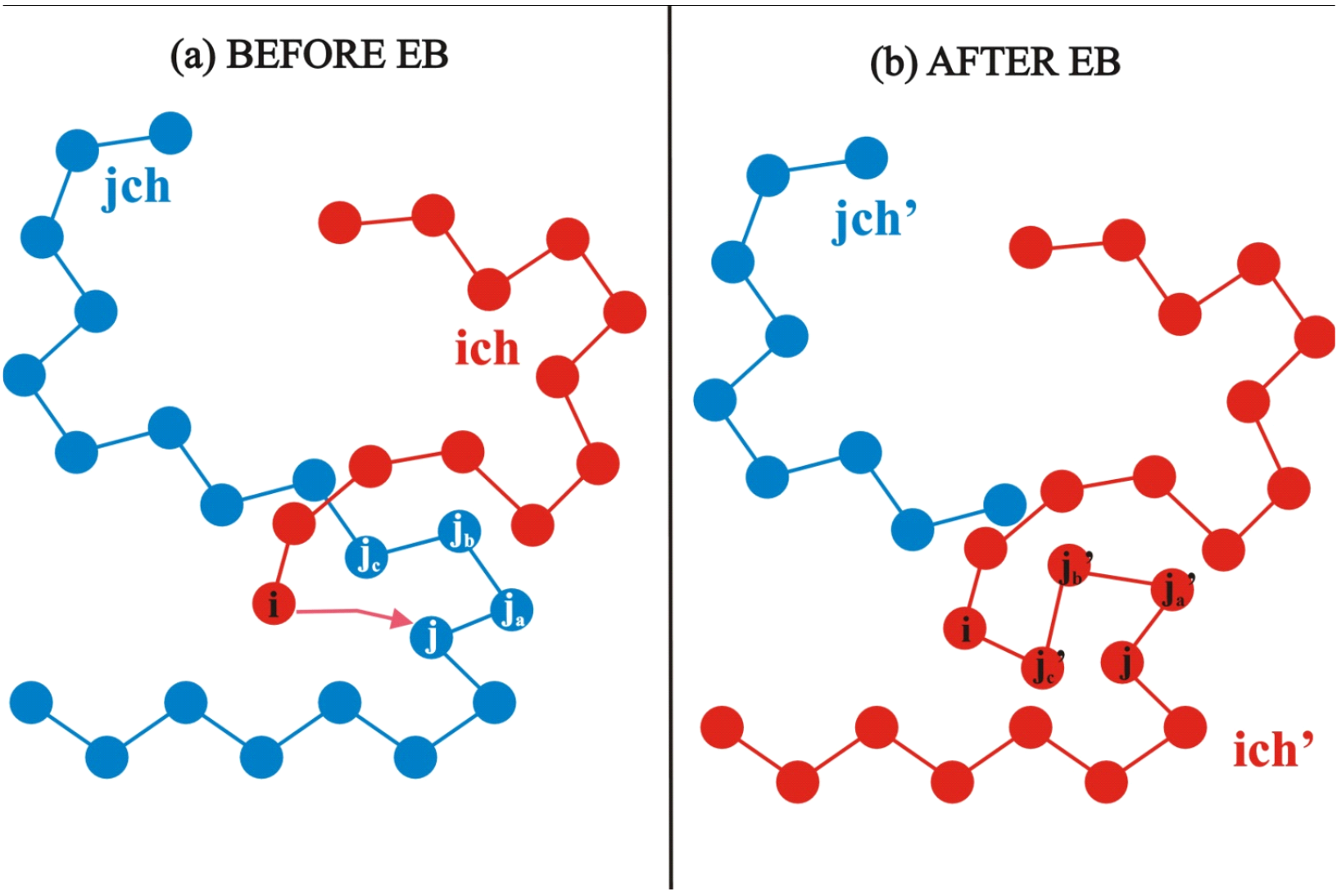}}
  \caption{(color online) Schematic of the end-bridging (EB) move. Left: initial configuration. The red arrow indicates that atom $i$ of chain $ich$ will attack atom $j$ of chain $jch$;
   then, trimer $(ja, jb, jc)$ will be excised from the system. Right: chain configurations after the application of the move. Atoms $i$ and $j$ are connected through the
   new trimer bridge $(ja', jb', jc')$. Reproduced from \protect\cite{Karayiannis2006}. \label{fig_endbridging}}
\end{figure}

The latter method is similar in spirit to the end bridging (EB) Monte Carlo method, see Fig.~\ref{fig_endbridging}, which has been developed extensively
in the groups of Mavrantzas and
Theodorou \cite{Mavrantzas1999,Harmandaris2000,Doxastakis2001a,Doxastakis2001b,Uhlherr2001,Uhlherr2002,Karayiannis2002,Karayiannis2006,Baig2007,Baig2009,Karayiannis2009}.
A thorough analysis of the geometric formulation and numerical implementation of the original EB algorithm is given in Ref. \cite{Mavrantzas1999}.
Using this method, equilibrated samples of polyethylene (up to C$_{500}$) \cite{Mavrantzas1999} and cis-1,4-polyisoprene (up to C$_{200}$) \cite{Doxastakis2001a,Doxastakis2001b}
have been generated. Later faster methods were developed with directed moves, dubbed Directed Internal Bridging (DIB) and Directed End Bridging (DEB),
which enabled them to equilibrate polyethylene samples as long as C$_{6000}$ \cite{Uhlherr2001,Uhlherr2002}. Extensions of the end-bridging
algorithm to different chain architectures have also been introduced \cite{Karayiannis2002}.

An exciting development, of relevance to the rheology of polymer melts, is the use of end bridging moves in combination with a small tensorial field to obtain
well-equilibrated, pre-oriented, strained configurations \cite{Harmandaris2000,Baig2007,Baig2009}. This way, viscoelastic properties can be measured
from atomistically detailed models, although up to now only simulations with unentangled chains have been reported.


In light of section \ref{sec_reinserting} we already mention that atomistic conformations can also be equilibrated efficiently by mapping them onto
a coarse-grained model, equilibrating the coarse-grained model, and finally reinserting the atomistic details.
Of course End Bridging moves can also be applied in these coarse-grained equilibration runs \cite{Kamio2007,Mulder2008a,Mulder2008b}. 


\subsection{Examples of atomistic simulations of dynamic properties}

Given the very steep increase of terminal relaxation times with molecular weight, it should come as no surprise that atomistically detailed MD
simulations of the dynamics of polymer melts have only been performed for relatively short unentangled or slightly entangled chains.

Polyethylene, because of its simplicity, has been one of the most favourite polymers studied in the past.
Starting in 1995, Paul and co-workers studied the diffusion coefficient of unentangled chains,
C$_{44}$ \cite{Paul1995} and C$_{100}$ \cite{Paul1997}, both by an explicit atom model and by a united atom model.
In this united atom model, which was optimised to reproduce experimental $P$-$V$-$T$ behaviour, each methyl CH$_x$ group is treated as one particle.
The viscosity of C$_{100}$ was investigated by Moore et al. \cite{Moore2000}.
Later, Harmandaris et al. studied the diffusion coefficient of samples of C$_{24}$, C$_{78}$ and C$_{156}$ polyethylene which had been
previously relaxed by the EB algorithm \cite{Harmandaris1998}. The shear relaxation modulus $G(t)$ of the C$_{24}$ and C$_{78}$ samples
were also obtained by analysing equilibrium stress fluctuations \cite{Harmandaris2000}. Soon after, in 2001, our group reported 
the mean square displacement, shear relaxation modulus, and viscosity of a C$_{120}$ polyethylene sample \cite{Padding2001}. Finally, in 2003
mean square displacements and diffusion coefficients for chain lengths C$_{78}$ to C$_{250}$ were reported \cite{Harmandaris2003}.

Other polymer species have been studied as well, notably 1,4-polybutadiene. Whereas Smith and co-workers in 1999 were still limited to 100 carbon
atom simulations \cite{Smith1999}, in 2005 Tsolou and co-workers were able to study the self-diffusion and dynamic structure factor of polybutadienes 
with up to 400 carbon atoms \cite{Tsolou2005}.

Despite the fast growth in computing power, the extremely fast increase of the longest relaxation time with chain length, $n^{3.4}$, implies that
atomistically detailed MD simulations will remain limited to only slightly entangled samples. Longer chains can and have been studied, for example
Ryckaert studied a melt of fully atomistic C$_{1000}$ chains \cite{Ryckaert2005}, but this can only be done for simulation times much shorter than the longest
relaxation time.


\section{Coarse-graining: lumping together the atoms}
\label{sec_theory}

\subsection{Theory of coarse-graining}

In order to increase the time and length scales accessible in the simulation of polymers,
detailed atomistic models are replaced by coarse-grained models in which each particle
represents a collection of atomic particles.
In this review we focus on the bottom-up (\textit{ab initio}) approach where the interactions
between the coarse-grained particles are directly linked to the atomistic interactions and are
aimed at correctly reproducing the structural, thermodynamic and/or dynamical properties
\cite{Akkermans2000,Baschnagel2000,Muller-Plathe2002,Paul2004,Padding2007}.

When coarse-graining polymers, the polymer chain is subdivided into subchains, partitioning the
degrees of freedom in two sets: the coarse-grain coordinates and momenta $\{R,P\}$, which
are the centres of mass (or characteristic atom) positions and momenta of the subchains, and the `bath' coordinates
and momenta $\{q,p\}$, which are the remaining internal coordinates and momenta describing the details of the configurations.
The potential of mean force governing the forces among the coarse-grained coordinates is given by
\begin{equation}
A(R) = -k_BT \ln \int \mathrm{d} p \ \mathrm{d} q \ \exp \left( - \beta H_B(p,q;R) \right) \mbox{,} \label{eq_potmeanforce}
\end{equation}
where $\beta = 1/(k_BT)$ and $H_B(p,q;R)$ is the bath Hamiltonian, equal to the sum of kinetic energy $T_B(p,q)$ of the bath variables
and the potential energy $\Phi(R,q)$ of the entire system. In a system containing only
coarse-grain particles, employing the potential of mean force $A(R)$ ensures a correct distribution of $R$ coordinates,
as well as correct thermodynamic properties. Unfortunately, the potential of mean force
is generally a complicated function of \textit{all} coordinates $R$, and invariably includes
complicated multi-body interactions. For practical and computational reasons it is impossible to
calculate and store the potential of mean force for all possible multibody configurations. Some approximations
need to be made. The most widely used approximation is that the potential of mean force can be represented by
pairwise and sometimes triplet terms.  We would like to issue a warning that a naive use of the interactions
at the coarse-grained level can lead to incorrect values for some \textit{thermodynamic} properties.
More specifically, the thermodynamics of the coarse-grained system will not correspond to the
thermodynamics of the microscopic system if one falsely assumes that one is still dealing with an
atomistic system, where the coarse-grain particles are the only particles present and the effective pair
interactions are the only energy terms present. It should be intuitively clear that the relatively softer
interactions will then lead to a pressure which is generally too small and
a system which is too compressible. The source of this thermodynamic
inconsistency can be traced to the state point dependency as well
as the pair approximation of the potential of mean force.
For a discussion on these matters, the reader is referred to \cite{Briels2002,Likos2001,Louis2002}.

The theory underlying coarse-graining of \textit{dynamics} is the Mori-Zwanzig formalism \cite{Zwanzig1961,Akkermans2000,Kinjo2007,Padding2007},
where a projection operator technique leads to a Generalized Langevin Equation governing the dynamics of the coarse variables. Although the dynamics of the atomistic model is deterministic and conservative,
the dynamics of the Generalized Langevin Equation is stochastic and includes dissipation and
memory effects:
\begin{equation}
	\frac{\mathrm{d} P_n}{\mathrm{d} t}(t) = -\frac{\partial A}{\partial R_n}(t) + F_{n,t}^R -
\sum_m \int_0^t \mathrm{d} \tau \ P_m(t-\tau) \zeta_{mn}\left( \tau;R(t-\tau) \right). \label{eq_gle}
\end{equation}
Here $P_n$ is the momentum of coarse-grained particle $n$, $A$ is the above-mentioned potential of mean 
force, $F^R_{n,t}$ is a random force, and $\zeta_{mn}$ is a time-dependent friction matrix.
The Mori-Zwanzig formalism provides an exact derivation for this friction matrix, but its interpretation is not 
straightforward: $\zeta_{mn}\left( \tau;R(t-\tau)\right)$ 
is the correlation,
at the point where the coordinates $R$ equal $R(t-\tau)$, of the random force at time zero on particle $m$ 
with the random force propagated by a complicated operator on particle $n$. Specifically, this operator is
$\exp \{ (1-{\cal P}) i {\cal L} \tau \}$, where ${\cal L}$ is the Liouville operator and ${\cal P}$ projects onto
the coarse-grain coordinates \cite{Padding2007}.
If the theory is followed to the rule (including the exact potential of mean force), a correct
description of the structure and dynamics follows automatically. However, as the above equation shows,
the resulting friction on a particle again depends on 
\textit{all} coordinates $R$ and momenta $P$. Compared to the case of the potential of mean force the situation is aggravated,
because the friction depends also on the coordinates and momenta in the past. Clearly, for 
practical and computational reasons approximations need to be made. For example it is often assumed that the
friction is pairwise additive with the contribution of a pair depending only on the positions and momenta of the two particles
involved. Also memory effects are often ignored. This may or may not be correct. Making the right approximations, often
based on physical intuition, is the responsibility of the coarse-grainer.

\subsection{Classification of coarse-grained dynamics methods}

The term `coarse-grained dynamics' is used rather loosely in the literature. Many different levels
of coarse-graining are encountered, with different approaches to treating the friction.
In our view it is important to distinguish between two main classes, which may be termed \textit{coarse-grained molecular dynamics}
and \textit{coarse-grained stochastic dynamics}, see Fig.~\ref{fig_CGlevels}. The difference is the following:
\begin{itemize}
	\item \textbf{Coarse-grained molecular dynamics} (CGMD) applies to simulations where a few atoms (up to one chemical repeat unit or 
	five backbone carbon atoms) are coarse-grained	to relatively hard beads. In this case the additional friction components are
	often not dominant, and in most cases are ignored altogether.	Ignoring the friction leads to too fast dynamics, which is beneficial
	for equilibration, but requires a rescaling of time when quantitative comparison with real dynamics and rheology is desired.
	\item \textbf{Coarse-grained stochastic dynamics} (CGSD) applies to simulations where many atoms (many chemical repeat units or
	more than 10 backbone carbon atoms) are coarse-grained to relatively soft beads or `blobs'.
	The friction and stochastic forces dominate the interactions, and cannot simply be ignored but must be included in the equations of motion. Often it is assumed
	that the stochastic forces are delta-correlated, i.e. have no memory. Unless countermeasures are taken, the soft interactions cannot prevent
	crossing of the bonds between the coarse-grained particles, leading to unrealistic dynamics for entangled polymers.
\end{itemize}
In the following sections we will treat both levels of coarse-graining.

\section{Lumping a few atoms into beads}
\label{sec_cgmd}

\begin{figure}[tp]
\centering
  \scalebox{0.8}{\includegraphics{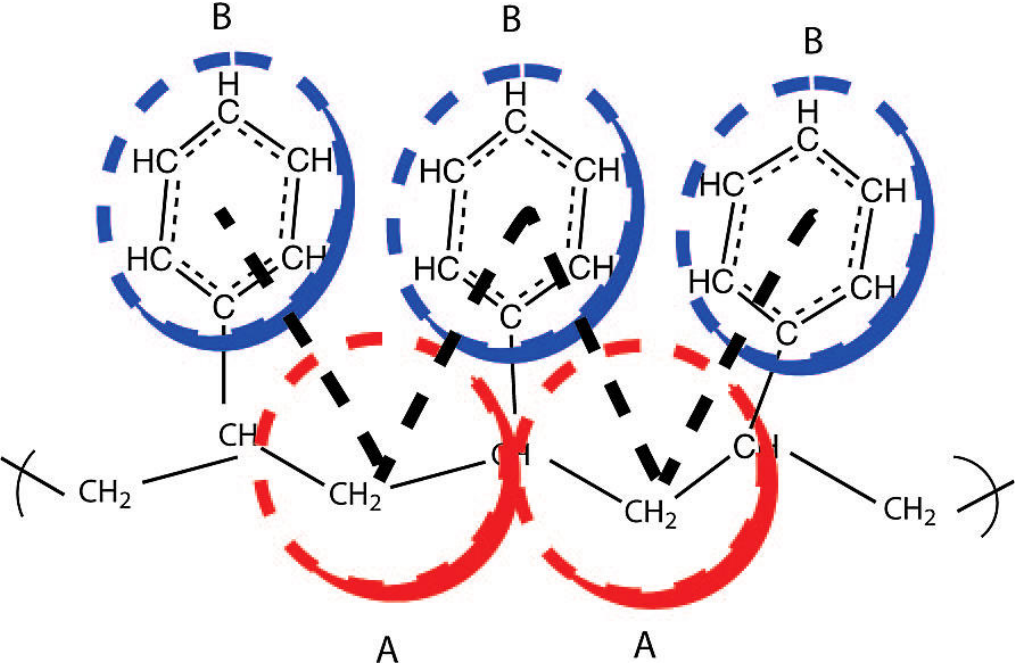}}
  \caption{(color online) Coarse-graining of polystyrene to two types of beads `A' and 'B' representing part of the backbone and a phenyl-ring, respectively.
  Dashed lines show the bonds between the coarse-grained beads A and B. Reproduced from \protect\cite{Harmandaris2009}. \label{fig_polystyrene}}
\end{figure}

Let us first focus on coarse-grained molecular dynamics (CGMD) simulations. A coarse-grained particle may represent the
centre-of-mass of a small group of atoms, or the position of a characteristic atom which is important in determining
the local structure of the polymer backbone or a sidegroup.
An example is given in Fig.~\ref{fig_polystyrene} where polystyrene chains are coarse-grained to `A' and `B' beads representing a part
of the backbone and the phenyl-ring, respectively \cite{Harmandaris2009}.

\subsection{Targeting the structure of real polymer chains}

Most coarse-graining efforts focus at reproducing the structure of the underlying microscopic chain \cite{Baschnagel2000,Muller-Plathe2002,Muller-Plathe2003,Faller2004,Aleman2009,Fritz2009}.

The structure of a polymer melt can be described in terms of distribution functions of distances and angles between
the coarse-grained particles. Once it is decided what the position of a coarse-grained particle represents,
it is rather straightforward to measure distribution functions for these coarse-grained coordinates in
atomistically detailed molecular dynamics simulations.
Intramolecular distribution functions include those for the distance between two adjacent coarse-grained beads, the angle
between three consecutive beads, and the dihedral angle between four consecutive beads. Other structural distribution 
functions include those for distances between beads on different chains, and can also include those for distances
between beads on the same chain but sufficiently far away along the backbone. 
These distribution functions are used as target functions for the coarse-grained interactions.

Usually, the local intramolecular interactions between consecutive beads are relatively strong and therefore rather unperturbed
by intermolecular interactions. In such a case a simple Boltzmann inversion is enough to obtain the coarse-grained interaction \cite{Faller2004}.
For example, from a distribution $P_{\mathrm{bond}}(r)$ of bond lengths $r$ between consecutive coarse-grained particles
(obtained in an atomistic simulation and already corrected for the Jacobian $4\pi r^2$), the coarse-grained bond potential is derived as
\begin{equation}
V_{\mathrm{bond}}(r) = -k_BT \ln P_{\mathrm{bond}}(r).
\end{equation}
One word of caution is at place here. Factorising the various distribution functions into independent parts representing bond distances,
bond angles and dihedral angles ignores any correlations which may be present between them \cite{Fukunaga2002,Harmandaris2007}.

\begin{figure}[tp]
\centering
  \scalebox{0.6}{\includegraphics{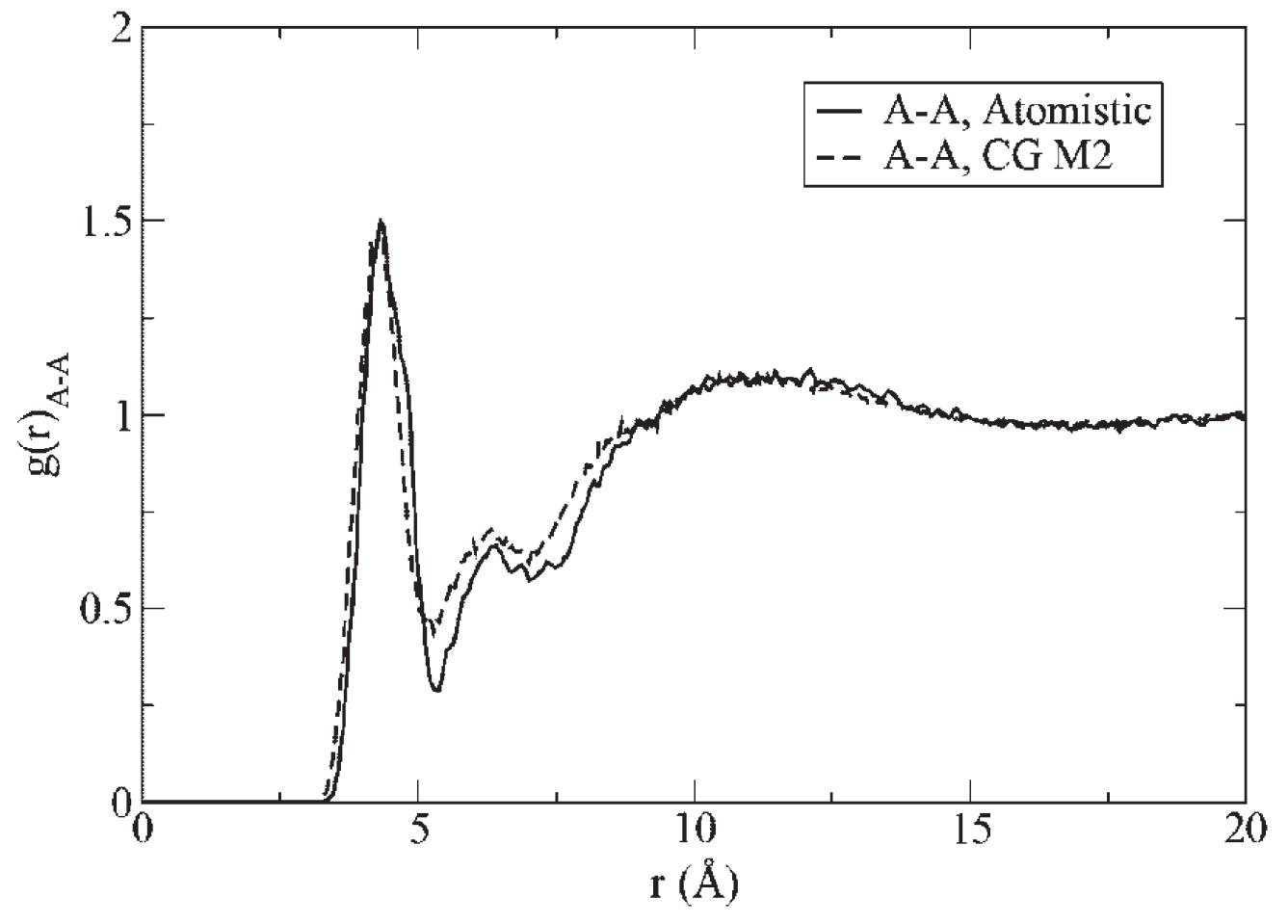}}
  \caption{Nonbonded A-A pair distribution function for a polystyrene melt (see Fig.~\ref{fig_polystyrene}) at $T=463$ K, obtained from atomistic
  molecular dynamics simulations (solid line) and coarse-grained simulation (dashed line). Reproduced from \protect\cite{Harmandaris2007}. \label{fig_grPS}}
\end{figure}

The nonbonded interaction, represented by the radial distribution function $g(r)$ between nonbonded coarse-grained beads, is however much more subtle.
It contains both entropic and enthalpic contributions and a simple Boltzmann inversion is insufficient because this ignores important
packing effects \cite{Fukunaga2001,Kremer2002,Faller2004,Fritz2009}. Different methods have been proposed to find the nonbonded interaction
which reproduces the target radial distribution function $g_{\mathrm{target}}(r)$.
One option is to choose a simple functional forms for the interaction, based on physical intuition,
and optimise the parameters of this function to reproduce the complicated radial distribution function as close as
possible \cite{Tschop1998a,Muller-Plathe2002,Muller-Plathe2003,Harmandaris2007}.
For example, the nonbonded interactions between the coarse-grained polystyrene beads of Fig.~\ref{fig_polystyrene} have been found by
optimising the parameters of a generalised Lennard-Jones-type interaction \cite{Harmandaris2007}. As Fig.~\ref{fig_grPS} shows, the
resulting nonbonded radial distribution functions agree very well with those obtained from atomistic simulations.

In some cases a predefined functional form is not flexible
enough to capture the intricacies of the target radial distribution function. More complicated analytical potentials,
defined piecewise for different ranges, allow more accurate reproduction of the complicated distribution
functions \cite{Reith2003,Tschop1998a,Abrams2003}. The parameters of these models may be obtained semi-automatic or
fully-automatic using simplex optimisation \cite{Faller2004,Meyer2000,Reith2001,Sun2005,Zhao2005}.

Another popular technique to obtain the coarse-grained interactions is Iterative Boltzmann Inversion (IBI) \cite{Reith2003,Faller2003,Faller2004,Sun2006a}. 
Focusing on the non-bonded interactions, for every iteration a one-to-one correspondence between the effects at a distance $r_0$ and the
non-bonded potential $V(r_0)$ at the same distance is assumed. In the beginning, a starting potential $V_0(r)$ has to be guessed,
which is usually the Boltzmann inversion of the target distribution function. At each iteration stage $n$, a simulation is performed with
the last guess $V_n(r)$ of the non-bonded potential, and the radial distribution function $g_n(r)$ is measured.
This leads to a new guess for the non-bonded potential:
\begin{equation}
V_{n+1}(r) = V_n(r) + k_BT \ln \frac{g_n(r)}{g_{\mathrm{target}}(r)}.
\end{equation}
The iterations continue until the difference between $g_n(r)$ and $g_{\mathrm{target}}$ is deemed satisfactory. Usually only a few iterations
are necessary \cite{Reith2003,Faller2004}.

Other techniques have been used as well. Within our own group we have decribed coarse-grained interactions by a number of
functions with coefficients which are left for automatic optimisation by treating them as dynamical parameters in an extended
ensemble \cite{Akkermans2001}.
Notably, Ashbaugh and co-workers used a combination of molecular dynamics simulations and inverse Monte
Carlo methods to self-consistently map structural correlations from atomistic simulations of alkane oligomers onto
coarse-grained potentials \cite{Ashbaugh2005}.

A warning is at place here. All techniques described above aim at reproducing the structure by means of pair interactions. This means
that the true potential of mean force, Eq.~(\ref{eq_potmeanforce}), is only approximated and many-body interactions beyond a certain
number of particles (usually two) are ignored. As a consequence, it is not guaranteed that the thermodynamic state is correctly
described \cite{Akkermans2001,Reith2003,Faller2004}.

An option is to include thermodynamic properties in the optimisation scheme. For example Reith et al. \cite{Reith2003} added an extra
pair-potential with an absolute value decreasing linearly to zero at a cut-off range. The amplitude (positive or negative) and cut-off
range were used as new parameters for re-optimisation. This yielded a radial distribution function which did not deteriorate too strongly,
combined with a correct pressure. Also, Nielsen et al. \cite{Nielsen2003} have developed a coarse-grained model for n-alkanes (polyethylene)
by optimizing bond and bend parameters and the non-bonded Lennard-Jones parameters to match observables from atomistic simulations as
well as experimental surface tension and bulk density data.

As already mentioned, when coarse-graining only a few atoms to beads the interactions are still relatively strong. This means that,
even though longer simulation times can be reached, equilibration is still sometimes an issue. In such cases the equilibration techniques
developed for atomistically detailed models can be used with high efficiency. For example, recently End-Bridging techniques have been applied to 
coarse models of poly(ethylene terephtalate) and atactic polystyrene \cite{Kamio2007,Mulder2008a,Mulder2008b,Spyriouni2007}. 

In summary, when coarse-graining one should always keep an eye on what properties one is interested in. Because in practice one is
limited to approximating the potential of mean force in terms of pair interactions, it is impossible to correctly represent both
the structural \textit{and} thermodynamic properties. Improving the agreement with one will invariably deteriorate the agreement with the
other, so compromises need to be made.

\subsection{Lattice models}

Up to this point we have only considered molecular models in continuous space. Atomistic models have also been mapped to lattice models.
One of the most important lattice models for polymers is the Monte Carlo-based Bond Fluctuation (BF) model \cite{Carmesin1988,Wittmer1992,Tanaka2000}, where each super-atom
occupied eight sites on a cubic lattice. Bond distances and angles can vary between different discrete values. The advantage of this
approach is that the possible distances and angles can be chosen in such a way that bond crossing becomes impossible. Mapping of real
polymers onto the BF model has been tried for bisphenol polycarbonates and polyethylene \cite{Baschnagel1991,Paul1991,Paul1994,Tries1997}.
Coarse-graining using the BF model is extensively reviewed in \cite{Baschnagel2000}.

Mattice and coworkers have developed a mapping from a rotational isomeric state (RIS) description of a polymer onto a diamond lattice for the
second nearest neighbour positions of polyolefine backbones \cite{Rapold1995,Rapold1996,Doruker1997,Haliloglu1998}. The intermolecular potential
in this model, with different values of the energy for different ranges of distances between the superatoms, was tuned by aiming at producing a
zero second virial coefficient (corresponding to theta-conditions) as well as a correct radius of gyration for polymers such as polyethylene
and polypropylene. For the dynamics local Monte Carlo moves, including a crankshaft move, were introduced.
Recently the dynamical updates have been redesigned to study the mean-square-displacement in polyethylene melts \cite{Lin2006} and in
bidisperse polyethylene melts \cite{Lin2007}.

\subsection{Backmapping: reinserting atomistic details}
\label{sec_reinserting}

An important task of CGMD models is to quickly generate well-equilibrated atomistic polymer structures. For this it is of course necessary
to be able to map an equilibrated coarse-grained model back onto an atomistic model. Such a backmapping procedure is possible, as has been
extensively reviewed in \cite{Baschnagel2000,Muller-Plathe2002}.

Among the first polymers to be equilibrated in this way are various polycarbonates \cite{Tschop1998a,Tschop1998b,Abrams2003,Hess2006},
but more recently backmapping procedures have also been applied to polystyrene \cite{Harmandaris2006,Harmandaris2007,Milano2005,Santangelo2007,Spyriouni2007}
and polyamide \cite{Carbone2008,Karimi-Varzaneh2008} chains.

An exciting new development, with possible applications to determination of the rheology, is the extension of backmapping to non-equilibrium situations \cite{Chen2009}.
In this method deformed conformations are maintained during backmapping by applying position restraints. The method has been demonstrated for 
atactic polystyrene under steady shear flow.

\subsection{Dynamic properties from coarse-grained molecular dynamics: rescaling of time}

CGMD simulations have not only been used to accelerate the equilibration of polymer melts, but also to study their dynamics \cite{Muller-Plathe2002,Paul2004}.
We have already discussed that additional friction factors and stochastic forces are inexorably linked to the procedure of coarse-graining.
However when the degree of coarse-graining is not too large, friction is often ignored, leading to too fast dynamics.
This may optimistically be referred to as `speed-up'. In such cases, when dynamic properties are presented,
it is assumed that all important processes are accelerated by the same factor (usually 2 to 5), i.e. that a simple
rescaling of time by a single scaling factor can recover the real dynamics. Although in most cases this assumption has no theoretical justification,
it nevertheless seems to be correct for low degrees of coarse-graining, as may be witnessed by the many successful applications of
time rescaling that may be found in the literature \cite{Hahn2001,Faller2002,Faller2003,Leon2005,Depa2005,Milano2005,Depa2007,Sun2006b,Harmandaris2006,Harmandaris2007,Harmandaris2009,Chen2006,Chen2008,Strauch2009}.

\begin{figure}[tp]
\centering
  \scalebox{0.65}{\includegraphics{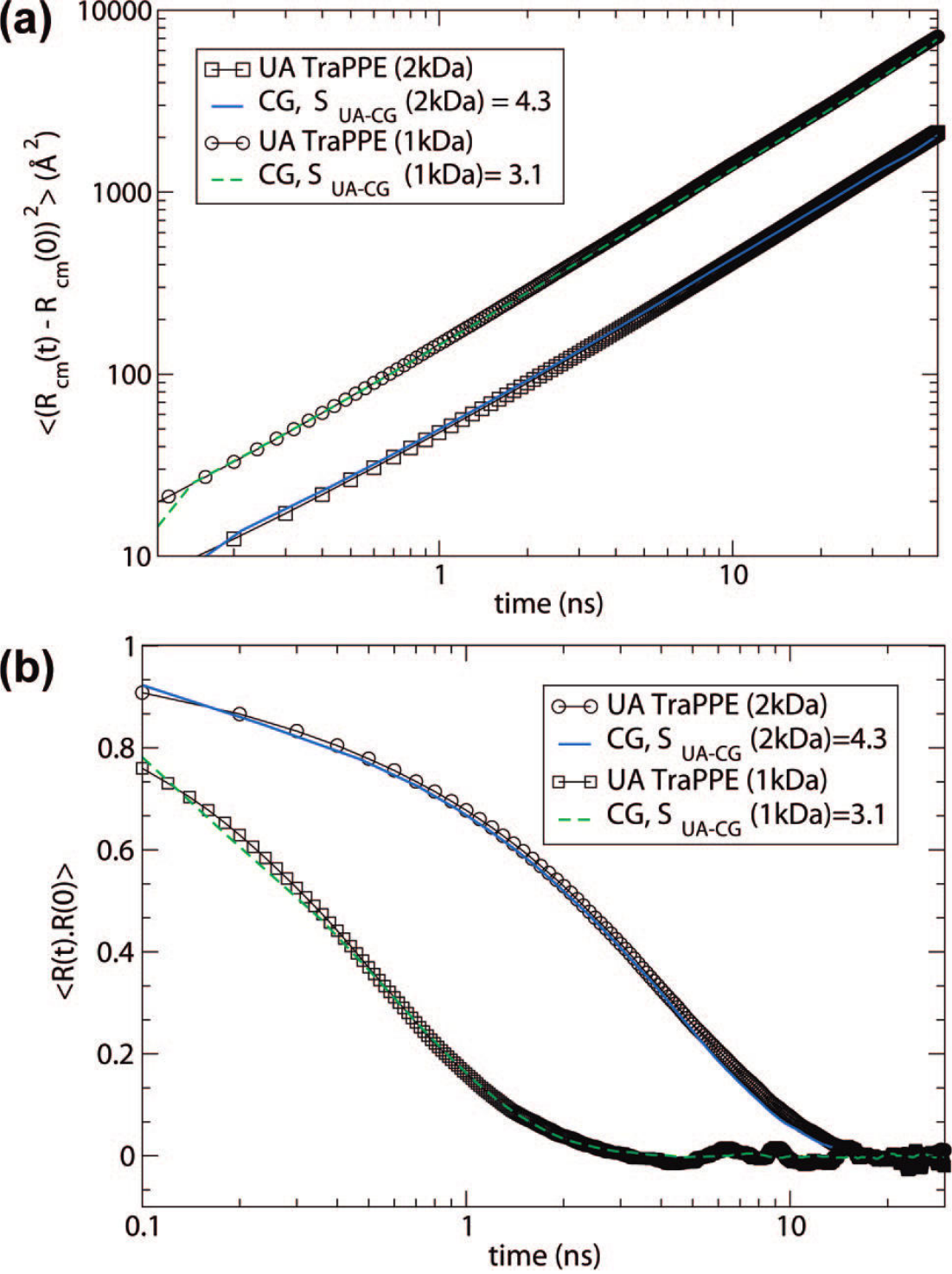}}
  \caption{Example of a successful time mapping of coarse-grained (CG) simulations onto chemically realistic united-atom (UA) molecular dynamics simulations of
  polystyrene. Two melts, with molecular weights 1 and 2 kDa have been studied, both at $T=463$ K. The time scaling factor $S$ needed for the chain centre-of-mass 
  mean-square-displacement (a) is consistent with the time scaling factor for the time correlation of the end-to-end vector (b).
  Reproduced from \protect\cite{Harmandaris2009}. \label{fig_PSmsd}}
\end{figure}

Calibrating the time-scale may be done by comparing chain diffusion coefficients,
although for most realistic polymers the atomistic simulations do not reach long enough times to observe free chain diffusion.
Since the coarse-graining level is relatively modest, there is hardly any difference between a bead in a relatively short (but still polymeric) chain
and a bead in a very long chain. Thus it can be expected that a diffusion-based calibration may be done using relatively short chains.
Harmandaris and co-workers have been very successful in
consistently determining the time scaling factors by comparing the atomistic and coarse-grained mean-square-displacements
of relatively short chains of polystyrene \cite{Harmandaris2006,Harmandaris2007,Harmandaris2009}.
Fig.~\ref{fig_PSmsd} shows a recent example \cite{Harmandaris2009}, where it is found that the mean-square displacement of the chain
centre-of-mass and the end-to-end vector time correlation can both be accurately reproduced by the same scaling factor $S$.
It is also possible to use other shorter local timescales to calibrate the time-scale. For example one can use the time associated
with the relaxation of subchains, as sampled by time correlations of the higher Rouse modes \cite{DoiEdwards}.

\section{Lumping many atoms into blobs: the role of friction and uncrossability}
\label{sec_cgsd}

In order to reach even larger time and length scales one may combine many more atoms, say $\ge 10$ monomeric units, into one coarse-grained particle.
The particle positions must be updated according to a coarse-grained stochastic dynamics (CGSD) scheme because the friction and stochastic forces often
dominate the interactions.
Neglect of friction and stochastic forces can lead to erroneous dynamics \cite{Oettinger2007,Qian2009}.
For example, the motion at certain length scales may appear as oscillatory, whereas the real polymer is overdamped at these scales.

\subsection{Friction and stochastic forces}

As already mentioned, the Mori-Zwanzig formalism of coarse-graining the dynamics leads to a Generalized Langevin Equation (GLE), Eq.~(\ref{eq_gle}), in which the friction and 
stochastic forces appear with memory \cite{Zwanzig1961,Kinjo2007,Padding2007,Guenza2003,Guenza2008,Hijon2010}.
Even though this equation is exact, it is very difficult
to use in practice. In the first place, evaluation of the memory terms requires taking averages over a projected dynamics which one does
not know exactly how to generate. In the second place, integrating the resulting integro-differential equations is an extremely challenging
numerical task.

Usually one therefore assumes that the time scale associated with changes in the positions of the coarse-grained coordinates is much
larger than the time scale associated with the decay of the friction memory. In this so-called Markov approximation the time dependence
of the friction $\zeta_{mn}$ in Eq.~(\ref{eq_gle}) is represented by a delta-function, resulting in an ordinary Langevin equation for the dynamics of the coarse-grained particles.
Whether or not the Markov approximation is valid must be carefully evaluated in each particular case.

Another difficulty remains: the friction and stochastic forces on a particle still depend on the positions and velocities of \textit{all} particles.
As already mentioned, it is often assumed that the friction is pairwise additive with the contribution of a pair depending only on the distance and relative 
velocities between the two particles involved. Doing so, we arrive at the pair friction model employed in the so-called dissipative particle dynamics (DPD)
method \cite{Hoogerbrugge1992}. Although this method is still relatively
costly, especially when the frictions are high, it is rather popular since it conserves momentum, which is a prerequisite for hydrodynamic
behaviour, and it is easy to use in non-equilibrium simulations. Within our group we have studied in detail the pair friction parallel and perpendicular to the 
axis connecting the centres-of-mass of two halves of a polymer chain in a melt \cite{Akkermans2000}.
Also, coarse-grained forcefields for polyethylene and cis-polybutadiene for use in DPD have been determined in the group of
Rousseau \cite{Guerrault2004,Lahmar2007}.
The Rousseau group discussed the dependence of the friction coefficient on the coarse-graining level in view of the overall scaling of the
dynamical properties.

The use of pair frictions in DPD is necessarily linked to the use of a second order propagator. This implies that one must use low frictions
to reach long time scales (high frictions require too small time steps). The damage an artificially
lowered friction does to the rheological properties is often counterbalanced by using harder conservative
interactions \cite{Nikunen2007}. This often leads to unrealistic structure on the scale of the bead.

Fortunately, in the case of polymer melts it is often admissible to use a scalar friction (i.e. with a static background), because the
friction may be thought of as being caused by the motion of a (part of a) chain relative to the rest of the material, which to
a first approximation may be taken to be at rest. Propagation of a velocity field as in a normal liquid is highly improbable,
meaning that hydrodynamic interactions are screened \cite{DoiEdwards}. The positions of the 
coarse-grain particles can then be updated by a first order Brownian dynamics propagator, using much larger time steps than are
possible with a second order propagator. 
One should be aware, however, that such a first order Brownian dynamics propagator is not Galilean invariant.
Extra care should therefore be taken when simulating flow of a polymer melt. Often it is assumed that the particles flow
affinely with the imposed deformation, although in the case of shear flow this assumption may be
relaxed \cite{Padding2008,vandenNoort2008,Sprakel2009}.

\subsection{Uncrossability of bonds and entanglement}

The advantage of lumping together only a few atoms in CGMD models is that the resulting interactions are still sufficiently strong to prevent bond crossing.
In bead-spring models, such as the important FENE model \cite{Kremer1988,Kremer1990,Gao1995,Smith1995,Gao1996,Smith1996,Faller1999,Aoyagi2000,Putz2000,Kroger2004,Hou2010},
uncrossability is ensured by the use of relatively hard beads. Such models therefore belong to the class of CGMD models. Relatively small
friction and stochastic forces have been added to bead-spring models, but with the goal of stabilising the integration of the equations of motion
over very long time scales \cite{Kremer1990}, rather than representing the \textit{physical} friction caused by the degrees of freedom that have been coarse-grained out.

\begin{figure}[tpb]
\centering
  \scalebox{0.35}{\includegraphics{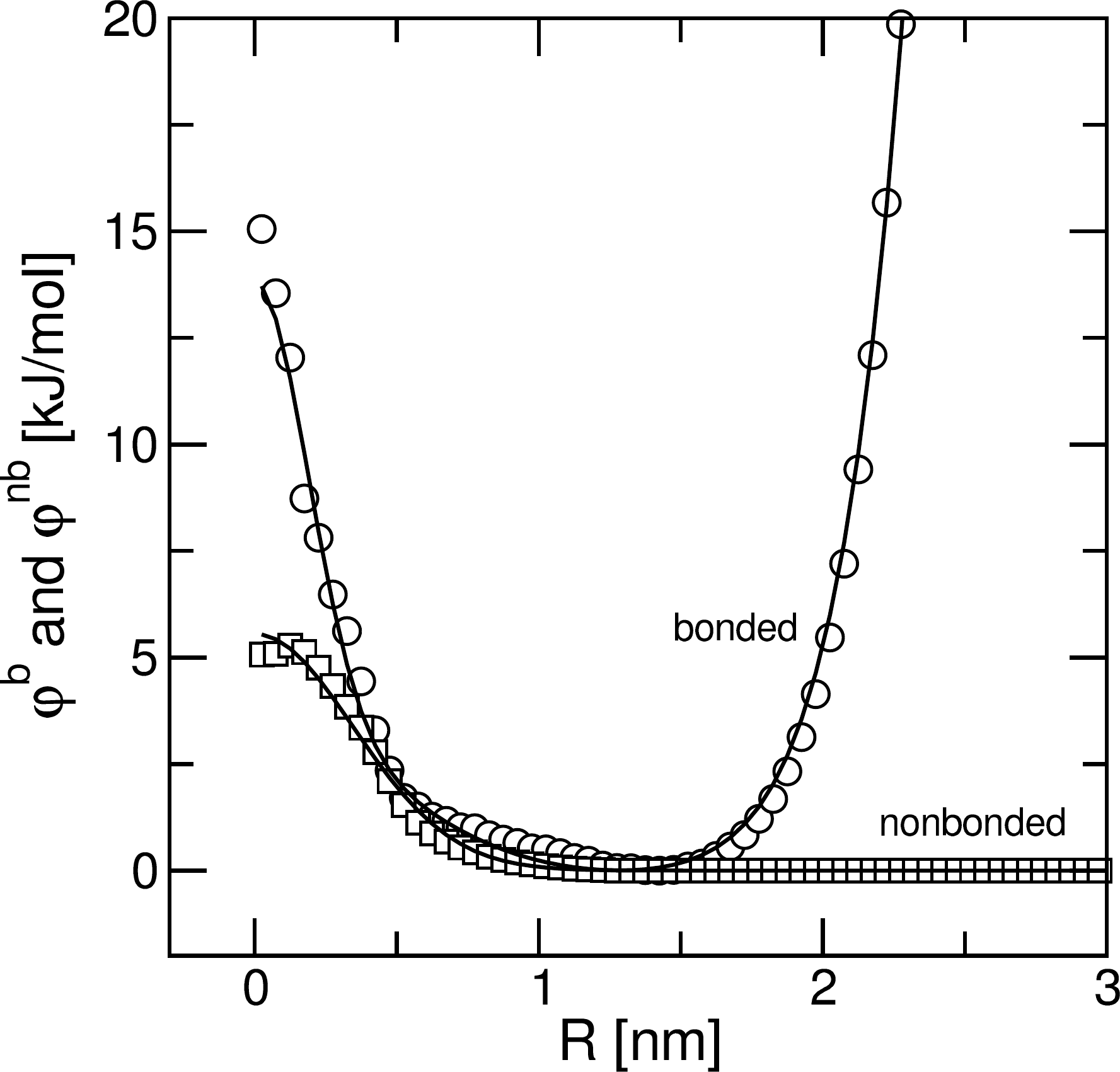}}
  \caption{The potential of mean force between bonded (circles) and nonbonded
  (squares) coarse-grained pieces of polyethylene ($T = 450$ K), each piece representing the centre-of-mass
  of 20 carbon groups. These potentials were obtained from distribution functions
  measured in atomistically detailed molecular dynamics simulations of C$_{120}$H$_{242}$. 
  Note that $k_BT = 3.74$ kJ/mol, meaning that without additional measures bonds can easily cross each other.
  \label{fig_potmeanforce}}
\end{figure}

As more and more atoms are coarse-grained into one particle, the interactions between them become softer and softer \cite{Padding2001a,Strauch2009,Lahmar2009},
see Fig.~\ref{fig_potmeanforce}.
Beyond a certain degree of coarse-graining, bonds will be able to cross each other. Without additional measures the important
entanglement effect, leading to altered and much slower dynamics, is therefore lost.

\begin{figure}[tpb]
\centering
  \scalebox{0.5}{\includegraphics{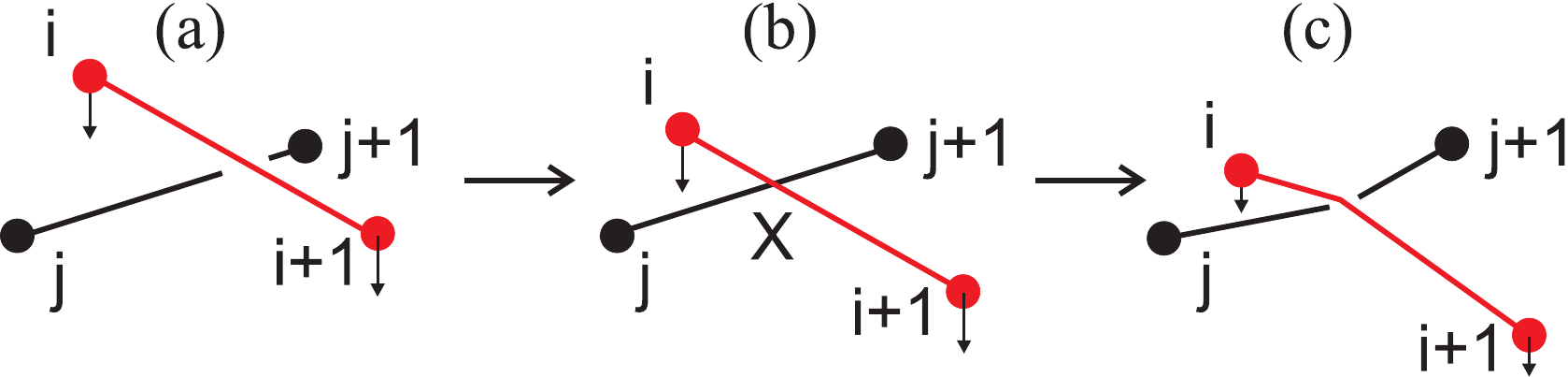}}
  \caption{(color online) 
  Principle of the Twentanglement algorithm. (\textit{a}) Two line segments representing a bond are
  closing in on each other. (\textit{b}) At a certain moment these bonds will touch. An
  `entanglement' is created at the crossing point $X$. (\textit{c}) After this, the bonds
  are viewed as slippery elastic bands. The elasticity will slow down the relative speed
  of the bonds. This sequence of events may also be reversed. \label{fig_twentanglement}}
\end{figure}

An algorithm that can detect and prevent unphysical bond crossings has been developed by us \cite{Padding2001a}.
Employing this so-called Twentanglement algorithm to a simulation of highly coarse-grained polymer chains
will reintroduce the entanglement effect. The principle of the algorithm is depicted in Fig.~\ref{fig_twentanglement}.
The bonds are considered to be elastic bands between the bonded particles. 
The algorithm keeps track of all (unattached) bond vectors which are close together.
For each bond vector and at each instant of time the following triple product is calculated:
\begin{equation}
	V_{ij} = \left( \mathbf{r}_i - \mathbf{r}_j \right) \cdot \left[ \left( \mathbf{r}_{i+1} - \mathbf{r}_i \right)
	\times \left( \mathbf{r}_{j+1} - \mathbf{r}_j \right) \right], \label{eq_triple}
\end{equation}
where we have used Fig.~\ref{fig_twentanglement} as a reference for the indices. The absolute value of
Eq.~(\ref{eq_triple}) is the volume of the parallelepiped defined by the vectors $\mathbf{r}_{i+1,i}$,
$\mathbf{r}_{j+1,j}$, and $\mathbf{r}_{i,j}$. Aside from some pathological cases \cite{Padding2001a}, if $V_{ij}$
changes sign from one time step to the next, a bond crossing may have occurred. Additional checks
are made to ensure that the crossing is taking place along the physical part of the line segments (the above
equation checks if two \textit{infinite} lines have crossed). If a real bond crossing has occurred, an
entanglement is created at the crossing point. Subsequently, the associated volume $V_{ij}$ will serve to
detect future disentanglements. If the volume $V_{ij}$ of the four objects surrounding an entanglement
changes sign, a possible disentanglement has occurred, i.e. Fig.~\ref{fig_twentanglement} may also
be read backwards. Usually only a few of the uncrossability constraints contribute to an entanglement in the usual sense of 
longlasting obstacles, slowing down the chain movement. For instance, a C$_{60}$H$_{122}$ chain is
generally considered not to be entangled, yet many `entanglements' occur in a coarse-grained
simulation. The dynamics of the `entanglement' points is determined by a balance of forces, as
described in detail in \cite{Padding2001a}.

\begin{figure}[tpb]
\centering
  \scalebox{0.35}{\includegraphics{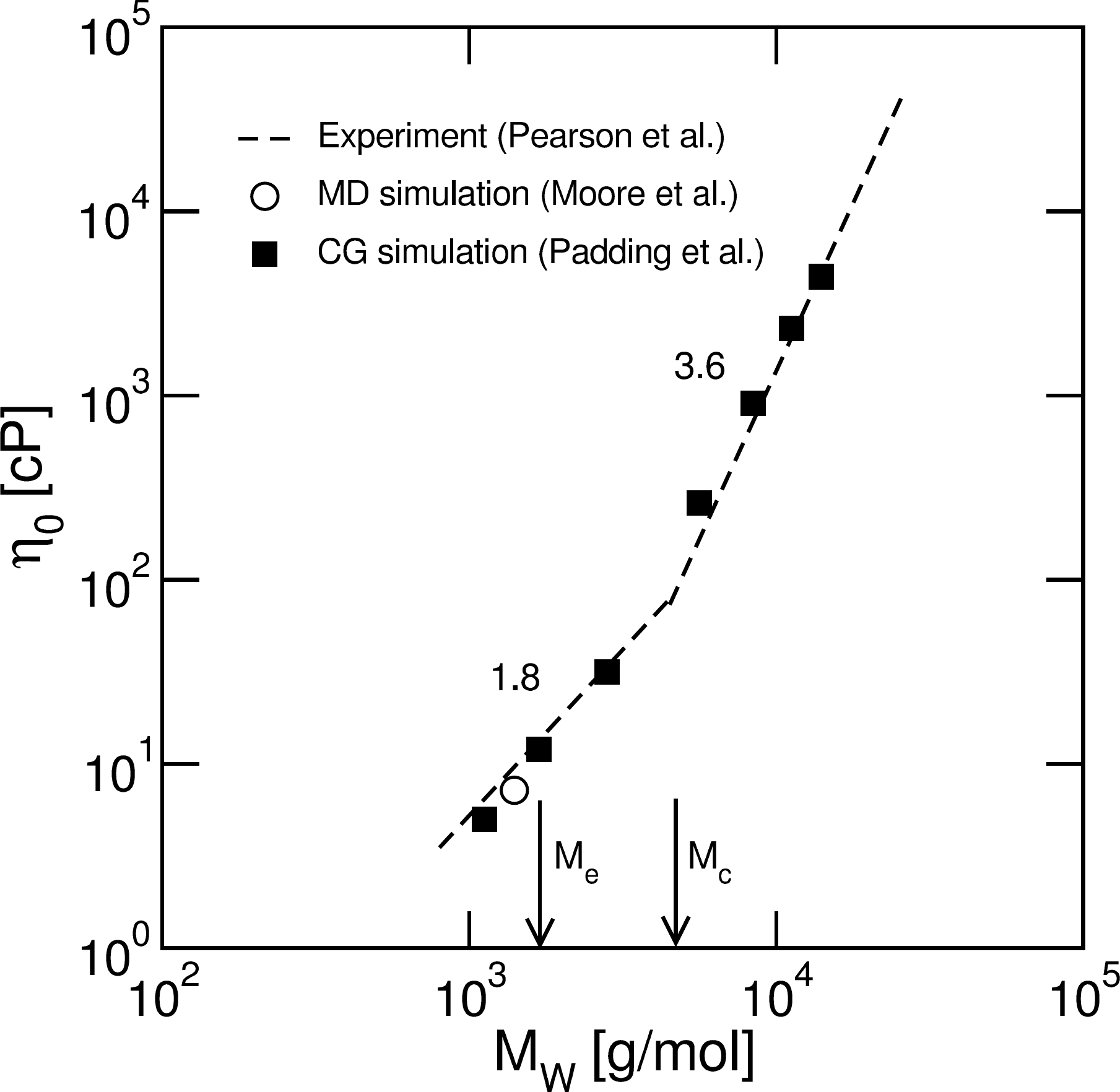}}
  \caption{The zero-shear viscosity $\eta_0$ as a function of molecular weight $M_w$ for polyethylene at $T=450$ K from coarse-grained
  simulations using Twentanglement \protect\cite{Padding2002} (solid squares) and experiment \protect\cite{Pearson1987} (dashed lines). The result from a
  chemically detailed simulation of C$_{100}$ \protect\cite{Moore2000} is also included (open circle). Observe that the steep increase in viscosity
  does not start at the entanglement molecular weight $M_e$ but at a critical molecular weight $M_c$ which is several times higher \protect\cite{Padding2002}.
   \label{fig_PEvisc}}
\end{figure}

Because uncrossability is explicitly taken care of, there is much freedom to choose the number of
monomers per coarse-grained particle. In our previous work, the following considerations were taken
into account: (1) the degree of coarse-graining should be large enough to allow for a significant
increase in the time and length scales accessible to simulation, and (2) if one wants to study reptation effects, the degree of coarse-graining
should not be so large that the size of a coarse-grained particle exceeds the typical diameter of the tube in the reptation
picture, in other words the entanglement length. A suitable choice for polyethylene was to coarse-grain
20 CH$_2$ units to one particle \cite{Padding2001a}. We have also been able to measure the required
friction on such a coarse-grained particle from a chemically realistic simulation by analysing the
correlations in the constraint force required to keep the location of a coarse-grained particle fixed
in an atomistically detailed simulation \cite{Padding2002}. By using the Twentanglement algoritm we have been able to predict the
dynamics and rheological properties of polyethylene melts up to C$_{1000}$ (Fig.~\ref{fig_PEvisc}) \cite{Padding2002,Padding2003,Padding2004,Kindt2005} and the dynamics of poly(ethylene-alt-propylene) melts \cite{Perez2010}.
We also applied the method to entangled wormlike micelles of a specific surfactant chemistry, and
found quantitative agreement with experimentally determined rheology \cite{Padding2008}.

Analysing the results of these coarse-grained simulations we found that the chain stiffness is an important ingredient for
the dynamics of polymer chains. Indeed, Faller and M\"uller-Plathe noted that chain stiffness intensifies the reptation
characteristics of polymer melt dynamics \cite{Faller2001,Faller2002}.

Topological constraints have also recently been introduced in DPD simulation methods \cite{Nikunen2007,Liu2008,Lahmar2009}. In standard DPD, the conservative
forces between the polymer segments decrease linearly with increasing pair distance. The interactions are therefore naturally quite soft and
polymer chains built from such soft beads behave as phantom chains who pass freely through each other \cite{Spenley2000}.
Nikunen \textit{et al.} prevented such chain crossing by simply increasing the strength of the DPD forces \cite{Nikunen2007},
whereas Liu \textit{et al.} added a rigid core to the DPD spheres \cite{Liu2008}. Note that both these schemes essentially
make the conservative interactions more rigid, which is opposite to the trend observed when coarse-graining to higher levels.
One of the dire consequences is the introduction of unwanted structure on the scale of the DPD particle, which makes these methods
unsuitable for high degrees of coarse-graining.

Uncrossability of chain bonds can alternatively be introduced by introduction of an additional repulsive interaction which is based on the distance
of closest approach between two bonds \cite{Kumar2001,Lahmar2009,Hoda2010}.
Another possibility to conserve the topology, at least for Brownian dynamics simulations,
is to simply forbid random displacements that lead to crossing of bonds \cite{Ramanathan2007}.
In a similar vein, one of us has very recently developed an algorithm which
efficienly treats collisions between infinitely thin bonds in such a way that momentum and energy are conserved locally \cite{Padding2009}.
This allows the study of the relative importance of topological and hydrodynamic interactions.
It should be noted that in all methods described in this paragraph the bonds themselves stay rigid. A collision between two such bonds is a
hard collision. In real polymer melts the bonds between consecutive coarse-grain beads are highly flexible, and the collisions will consequently
be much softer (e.g. as in Twentanglement). As noted above \cite{Faller2001,Faller2002}, the intrinsic stiffness of the bonds may lead to artifically
enhanced entanglement effects. These methods are therefore more suitable for simulation of semiflexible chains, or for flexible chains at such a
low degree of coarse-graining that local chain stiffness is still important.



\section{Primitive paths}
\label{sec_primitive}

We have seen in the previous section that beyond a certain level of coarse-graining it becomes necessary to somehow maintain the information about the
topology of the system, otherwise the important entanglement effect is lost. One option is to use algorithms such as Twentanglement.
Another option is to switch to a one-chain model based on
the reptation model \cite{DoiEdwards}. This however poses several problems. Reptation theory is still facing many questions
which cannot be answered within the model. For example, to quote a recent review of Likhtman \cite{Likhtman2009}, it is still not clear what is
the entanglement or tube field restricting the chain motion. What are the statistical properties of the tube -- is it semiflexible on the length
scale of the tube diameter? What is the microscopic basis for the assumptions used to describe constraint release in the linear and non-linear
regimes? What happens to the entanglements and the tubes after large deformations or in fast flows? A practical
question of relevance for quantitative predictions is how the numerical parameters of the tube model, such as the entanglement mass and tube
diameter, can be related to the chemical composition of the polymer.

\subsection{Primitive path analysis}

The questions posed above have provoked an outburst of activities which try to determine the primitive path and entanglement
characteristics from more detailed simulations \cite{Kremer1990,Harmandaris2003,Everaers2004,Kremer2005,Kroger2005,Tzoumanekas2006a,Tzoumanekas2006b,Spyriouni2007,Depa2007,Aleman2009,Karayiannis2009,Tzoumanekas2009,Hoy2009,Baig2010}.
All efforts are based on Edward's view \cite{DoiEdwards} of a primitive path as the shortest path remaining when one holds chain ends fixed,
while continuously reducing (shrinking) a chain's contour without violating uncrossability \cite{Tzoumanekas2006a}. The chain contours are reduced
simultaneously until topological constraints block further shrinkage.

\begin{figure}[tpb]
\centering
  \scalebox{0.8}{\includegraphics{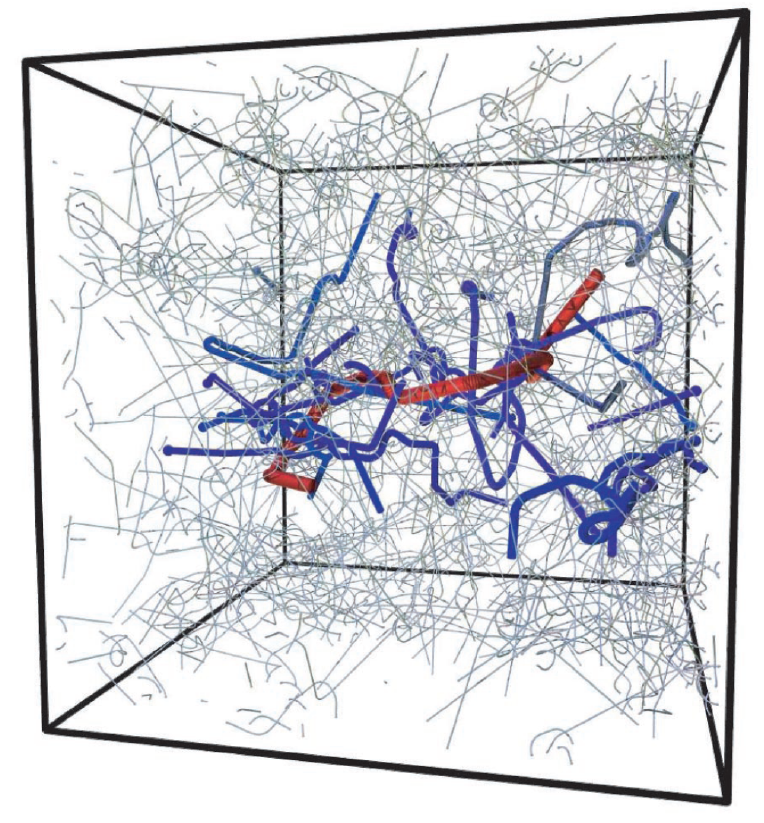}}
  \caption{(color online) Result of a primitive path analysis of a melt of 200 chains each consisting of 350 beads. The primitive path of one chain is shown (in red),
  together with all of those it is entangled with (in blue). The primitive paths of all other chains are shown as thin lines.
  Reproduced from \protect\cite{Everaers2004}.
  \label{fig_primitivepath}}
\end{figure}

Everaers \textit{et al.} introduced one such minimization method to extract the tube diameter and entanglement length from more detailed simulations
of linear chains \cite{Everaers2004,Kremer2005}, see Fig.~\ref{fig_primitivepath}.
On the basis of the tube model, both the real chain and the primitive path beyond some length scale are described
as random walk chains with the same mean square end-to-end distance $\left\langle R^2 \right\rangle$, but with different contour
lengths. The entanglement length and tube diameter $d$ correspond to the length and end-to-end distance of an entanglement strand, which is
identified with the Kuhn segment of the primitive path. Thus it is assumed that the following relations hold \cite{DoiEdwards}:
\begin{equation}
d = \frac{\left\langle R^2 \right\rangle}{\left \langle L_{pp} \right \rangle}, \qquad N_e = N  \frac{\left\langle R^2 \right\rangle}{\left\langle L_{pp} \right\rangle^2},
\end{equation}
where $\left \langle L_{pp} \right\rangle$ is the average primitive path contour length and $N$ is the number of Kuhn segments of the real chain.
$N_e$ is the entanglement molecular weight expressed as the number of Kuhn segments of the real chain.
A refinement of this method to exclude self-entanglement was introduced later \cite{Sukumaran2005}. It turned out that entanglements between
distant sections of the same chain make a negligible contributions to the tube and that the contour length between local self-knots is significantly larger
than the entanglement length.

The above method has been applied, among others, to polystyrene melts \cite{Harmandaris2009},
to semidilute solutions of stiffer chains such as actin \cite{Uchidaa2008}, and to bisphenol-A-polycarbonate \cite{Leon2005}, which was found to have
a surprisingly low entanglement length. The estimated entanglement molecular weight $M_e$ is usually in good agreement with the
experimental value obtained from the plateau modulus $G_0$.

The plateau modulus is a mechanical property of the system. Following a small step strain $\gamma_0$,
the shear stress $\gamma_0 G(t)$ initially decays according to local relaxation
processes on length scales smaller than the tube diameter. For well-entangled chains, after the entanglement time $\tau_e$, further relaxation of $G(t)$
is severely delayed by the topological interactions, leading to an apparent plateau;
only for times much longer than $\tau_e$ can $G(t)$ finally relax to zero.
The plateau modulus may be defined as the modulus at times larger than $\tau_e$, but much smaller than the terminal relaxation time.
The relation between the plateau modulus and the entanglement molecular weight is given by \cite{DoiEdwards}:
\begin{equation}
G_0 = \frac{4}{5} \frac{\rho RT}{M_e}. \label{eq_modulus}
\end{equation}
Here $\rho$ is the polymer (mass) density.
This expression is very similar (apart from the factor 4/5) to a prediction for the affine model of rubber elasticity \cite{DoiEdwards}.
Indeed, the good agreement between the entanglement length from the primitive path analysis and the entanglement length from the plateau modulus may intuitively
be understood because they both probe the same (albeit temporary) rubber-like elasticity of the underlying primitive network of chains.

The agreement with other estimates of the entanglement length is less good however, especially when these estimates are based on characteristic changes in 
the time-dependence of quantities which measure the motion of individual chains within their prospective tubes \cite{Tanaka2000,Padding2004}.
It may be argued that this disagreement is caused
by the dependence on non-exact theory in interpreting these measurements \cite{Tzoumanekas2006a,Hou2010}. This is certainly a factor, but the remarkable observation is that
with reasonable assumptions in most cases time-resolved measurements yield an entanglement length which is consistently larger than the one based on the plateau modulus.
For example for polyethylene, entanglement length estimates from the dynamic structure factor, the cross-over in Rouse mode time correlations, and the entanglement time $\tau_e$
(estimated directly from the time of the inflection point in $G(t)$) all give a value which is consistently larger by a factor of 1.5 compared to
the estimate from the plateau modulus \cite{Padding2004}.

The disagreement with the `plateau modulus' entanglement length may have two causes. First, dynamic
time-resolved quantities are probably more sensitive to details of the statistics (e.g. fluctuations) of the entanglements.
Second, dynamic time-resolved quantities
may be more sensitive to the finite stiffness of a real polymer chain on the scale of the tube diameter.

In a recent article \cite{Greco2008},
Greco noted that, because the tube diameter is relatively small, it is very important to include fluctuation effects in the description of entanglements.
Unfortunately, at the time of writing this review, there is still no definite agreement on the best way to determine the primitive paths and their
statistics.
Different coarse-graining schemes for arriving at the primitive path lead to differences in the statistics.
For example, Kr\"oger and co-workers proposed an alternative algorithm which returns a shortest path and related number of entanglements for a given configuration of polymers \cite{Kroger2005,Hoy2009}. Primitive paths are treated as infinitely thin and tensionless lines rather than multibead chains, and excluded volume is taken into account
without a force law. Their implementation allows construction of an optimal shortest path for 3D systems. The number of entanglements is then
obtained from the shortest path as either the number of interior kinks, or from the average length of a line segment.
This primitive path analysis has been applied to polyethylene C$_{24}$ to C$_{1000}$ \cite{Foteinopoulou2006,Foteinopoulou2009,Hoy2009},
as well as the FENE model \cite{Hoy2009}, validating analytical
predictions of Schieber \cite{Schieber2003a} about the shape of the distribution for the number of strands in a chain at equilibrium.
It was also concluded that the number of entanglements obtained by
assuming random walk statistics, as is done in the work of Everaers and co-workers, deviates significantly from the predictions by the algorithm of Kr\"oger and co-workers.

The random walk approximation partly fails because real polymers are typically still somewhat stiff at the scale of the average distance between two
uncrossability constraints \cite{Tzoumanekas2006a,Tzoumanekas2006b,Hoy2009,Tzoumanekas2009}.
As a result, the tube diameter cannot be equated to the Kuhn step length of the primitive path.
Although this fact is well-known for semiflexible fiber suspensions, where theoretical expressions for the tube diameter are different from expressions for the deflection length
(average distance between successive collisions of the fiber with its tube) \cite{Hinsch2007}, it seems to be often ignored in 
the polymer melt literature. Stiffness effects have been introduced recently in primitive path analyses by
Hoy \textit{et al.} \cite{Hoy2009} and by Tzoumanekas \textit{et al.} \cite{Tzoumanekas2009}.

\subsection{Primitive chain networks and slip-link models}

Although the precise characteristics of entanglements and/or confining tubes have remained somewhat elusive, this has not stopped researchers from
trying to predict polymer melt rheology by employing primitive path-based models.

\begin{figure}[tpb]
\centering
  \scalebox{0.7}{\includegraphics{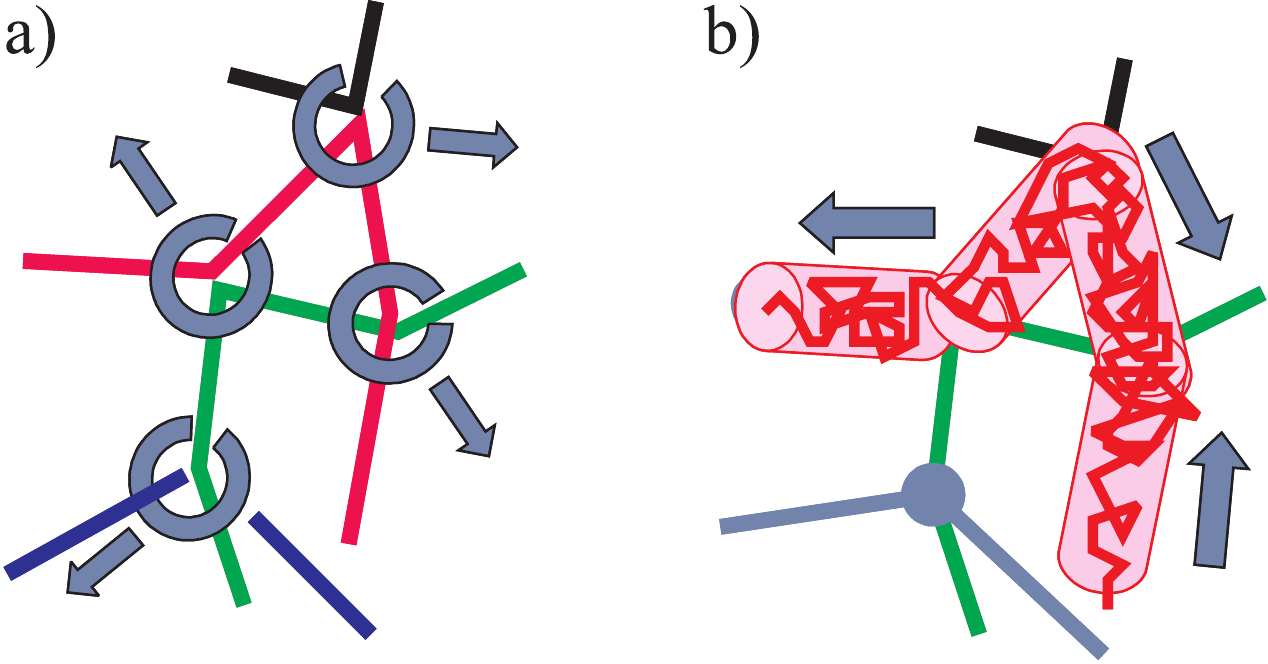}}
  \caption{(color online) Schematic representation of a primitive chain network model. (a) The motion of slip-links (rings) is influenced
  by the tension in the chain segments (arrows) and an osmotic force. (b) The motion of the monomers through the slip-links (arrows)
  results in reptative motion. (Picture kindly provided by Y. Masubuchi.)
  \label{fig_PCN}}
\end{figure}

An important development in this direction is the primitive chain (or slip-link) network model developed by Masubuchi and others \cite{Masubuchi2001,Masubuchi2003}.
In this model polymer chains are coarse-grained to the level of \textit{segments} between entanglements, see Fig.~\ref{fig_PCN}.
Chain coupling is achieved by confining two chains by a common slip-link placed at a specific position in 3D space. The chains fulfill
a force balance at the slip-links, similar to the treatment of uncrossability constraints in the Twentanglement
algorithm \cite{Padding2001a} which act on a somewhat smaller scale. The Langevin equation for the slip-links
contains both the tension in the chain segments emanating from the slip-link and an osmotic force
arising from density fluctuations. The motion of monomers through the slip-links ultimately generates reptation as
well as tube length fluctuations. When creating new slip-links (i.e. new entanglements),
a partner chain segment is chosen randomly among those spatially close to the advancing chain end.

The primitive chain network model correctly predicts the scaling of the longest relaxation time and the self-diffusion with chain length \cite{Masubuchi2001}.
The model was later modified to also include the concept of dynamic tube dilution \cite{Yaoita2004}, and good agreement between simulated and real linear and nonlinear rheology
was found. Extensions to branched polymers, blends and block copolymers have all been made \cite{Masubuchi2006a,Masubuchi2006b}.
Recently the model has been subjected to various sudden deformations to study the damping function, which is essentially the ratio of nonlinear stress relaxation
to the linear shear relaxation modulus \cite{Furuichi2007,Furuichi2008,Masubuchi2009}.
It was found that the force balance is a dominant correction over the basic Doi-Edwards theory as compared with the effect of convective constraint release.
Furthermore, the predicted normal stress ratio in shear, a quantity which is very sensitive to different modelling assumptions, was found to be in good agreement
with experiments.

In order to facilitate fast computations of the rheology, various single-chain slip-link models have also been
developed \cite{Tasaki2001,Doi2003,Shanbhag2004,Xu2006,Likhtman2005,Likhtman2007}. In such models an ensemble of primitive paths is followed in time.
Coupling between chains is not explicit, as in the above primitive chain network model,
but virtual by defining random relations between slip-links on different chains. Whenever a slip-link of a chain is destroyed due to tube renewal at
a chain end, the corresponding partner along the contour of another chain is also eliminated. The same applies to creation of new slip-links.

Very recently, the effectiveness of such
a single-chain slip-link model in describing the dynamics and rheology of entangled polymers was tested by systematically comparing slip-link
with FENE molecular dynamics results \cite{Sukumaran2009}.
The parameters of the model were determined by using the mean-square displacements of one particular chain length as a target function.
Using the same set of parameters, it was then tested if the predictions of the mean-square displacements for other chain lengths agreed with the MD calculations.
This was followed by a comparison of the shear relaxation moduli $G(t)$. The work identified a limitation of the original slip-spring model in describing the
static structure of the polymer chain as seen in MD, which was remedied by introducing a pairwise repulsive potential between the monomers in the chains. Also,
poor agreement of the mean-square monomer displacements at short times could be rectified by the use of generalized Langevin equations for the dynamics.
The latter of course somewhat diminishes the computational efficiency of the method.


Slip-link models are generally successful in their overall `rheological' performance and predictions \cite{Tzoumanekas2006a}. They also offer the possibility to
introduce new tube renewal mechanisms and, by turning these on or off at will, probe the effect on the rheology independent of other mechanisms.
A disadvantage of both single-chain and multi-chain slip-link models is that the link with the chemistry usually is far away; if a comparison with a microscopically
detailed model is made at all, the parameters are usually obtained by fitting rather than an ab-initio coarse-graining procedure.

An attempt at systematically bringing in structure from a detailed model, albeit again a
bead-spring type model, was introduced by Rakshit and Picu \cite{Rakshit2006}. In their method reptation is enforced because the chain inner blobs
are constrained to move along the backbone of the coarse grained chain (the primitive path), while the end blobs move in the three-dimensional embedding space.
These end blobs continuously redefine the diffusion path for the inner blobs. It was shown that various distribution functions, relaxation rates, and
the diffusion dynamics are properly represented. Recently, this model was used to study start-up and step strain shear flows \cite{Rakshit2008}. The authors showed that their
method reproduces several features observed experimentally such as the overshoot during start-up shear flow and shear thinning at large shear rates.

\section{Super coarse-graining: a polymer as a single soft particle}
\label{sec_super}

Industrial processing of polymers usually involves extrusion of a hot melt through a die or injection of a melt into a mould. Modern applications
also include nanoparticles in the melt to form composites. To optimise the process parameters for such complex flows it is important to understand
the interplay between time and length scales of the polymer and those imposed by the complex geometries.
The low frequency linear and non-linear rheology of polymer melts
is basically concerned with displacements of centres of mass only. Given the need to also reach large length scales, it is therefore just
natural to try to develop simulation models that only keep track of the dynamics of the centres of mass of all diffusing constituents.

An early application of this idea was presented by Murat and Kremer \cite{Murat1998}. These authors developed an efficient and rather general
model in which whole polymer chains are represented as soft ellipsoidal particles. The interaction between two such particles is taken to be
proportional to the spatial overlap of their monomer density distributions. Since the internal degrees of freedom of a chain are integrated out,
many thousands of chains can be simulated within reasonable computer time. More recently, McCarthy, Lyubimov and Guenza
introduced an approach based on the Ornstein-Zernike equations, in which a polymer is coarse-grained to a soft sphere on the scale of 
its radius of gyration \cite{McCarthy2009}. 
At such a high level of coarse-graining typical maximum interaction energies are of the order of $k_BT$, the thermal energy.

All these approaches focus on reproducing the static structure of a polymer melt, with good to
excellent agreement with the structure obtained from more detailed simulations.
However, because such models ignore memory and/or entanglement effects, they do not
have the characteristic slow dynamics of real polymer chains, and therefore cannot be used to predict the long-time rheology.

One may be tempted to slow down the dynamics of such a super-coarse-grained system by using a large friction term. 
Lyubimov \textit{et al.} recently introduced a method to estimate the friction on a polymer represented by a soft particle \cite{Lyubimov2010}.
Their approach is based on an intermediate mapping of a chemically detailed chain on a freely-rotating chain model and using a
hard-bead approximation to evaluate integrals involving the radial distribution function and dynamic structure factor. This yields a
series of diffusion coefficients in good agreement with results of atomistic simulations of unentangled and slightly entangled polyethylene melts.
No rheological predictions were made. 

The use of a single large friction term may be questioned, however, because 
in real entangled polymeric systems the frictions and random forces have memory of the configurations the system has gone through
in the recent and  sometimes even the distant past. A simple Brownian dynamics propagator with realistic mean forces and uncorrelated, fully random 
displacements without memory will not reproduce correct sequences of configurations of the retained coordinates.

In recent years, our group has introduced and explored a new efficient framework to reintroduce memory of previous configurations in
soft matter simulations \cite{Kindt2007,vandenNoort2008,Briels2009,Sprakel2009,Padding2010,Sprakel2011,Padding2011}.
In this framework, called Responsive Particle Dynamics (RaPiD), configurations and forces are described by as few variables
as possible. In case of polymer chains, similar to the works mentioned above, we usually restrict ourselves to three coordinates for
the location of each centre of mass.
To circumvent the complicated introduction of memory effects in friction forces and stochastic displacements
(as in the Generalized Langevin Equation, Eq.~(\ref{eq_gle})) we introduce a relatively
small set of additional dynamic variables, symbolically denoted as $n(t)$, which keep track of the thermodynamic state of the eliminated 
coordinates for the given values of the retained coordinates. If the configuration of retained coordinates is suddenly changed
(e.g. by a sudden change in the distance between the centres-of-mass of two polymers), the equilibrium values $n_0$ for the additional
variables change too, but the relaxation to $n_0$ takes place over a \textit{finite} time. This gives rise to strong transient forces
in addition to the thermodynamic forces deriving from the potential of mean force. 
It is possible to show that, to lowest order, the transient forces may be thought of as originating from a penalty free energy function which
is quadratic in the deviations of the additional variables \cite{Briels2009} .

The particular additional variables and corresponding names that are used to describe the transient forces depend on the system under study.
In some cases, such as for telechelic polymer networks, the variables are the number of bridging polymeric chains between nodes of the
network \cite{Sprakel2009,Sprakel2011}. In other cases, more relevant to this review on polymer melts, the variables are the number of
entanglements shared between a particular polymer chain and its immediate neighbours \cite{Kindt2007}.

To some readers it will be rather surprising that a single particle model would produce rheological results that are most naturally described by 
assuming curvilinear tubes surrounding each chain which prohibit the chains to move in any way other then along the primitive paths of the 
tubes. A few qualitative remarks may be of help at this point \cite{Briels2009}. First of all, the
main action of the tubes is to conserve the prevailing configurations of the chains and their centres of mass in particular.
This is exactly what the transient forces do as well. In a sense, the transient forces tie all centres of mass harmonically to their instantaneous 
positions. Secondly, beyond the Rouse time, \textit{i.e.} as soon as the chains perform one-dimensional diffusive motions along their tube-axes,
as a result of the Gaussian character of the tube-configurations the centres of mass of all chains perform simple random displacements. 
The only thing the tube does is to define the distribution of possible displacements and their averages in particular. Again this is what the 
transient forces do as well.

The RaPiD method has demonstrated its usefulness by correctly reproducing the large-time linear and non-linear rheology of linear
polymers \cite{Kindt2007}, telechelic polymers \cite{Sprakel2009,Sprakel2011}, star polymers \cite{Padding2010}, and polymeric
adhesives \cite{Padding2011}. The next challenge is to link the few additional parameters introduced
by the method to the chemical details of the system. In some cases, such as for the telechelic polymers \cite{Sprakel2009,Sprakel2011}, such a link has
already been successfully established.

\section{Conclusion}
\label{sec_concl}

The enormous range of time and length scales associated with the dynamics of entangled polymers precludes the use of one single computer 
simulation to calculate the large-scale rheological properties starting from the detailed chemical structure. Instead, some form of coarse-graining
is necessary. In this review we have made a distinction between coarse-grained molecular dynamics (CGMD) and coarse-grained 
stochastic dynamics (CGSD). 

In CGMD a few atoms (up to five backbone carbon atoms) are lumped into relatively hard beads. Uncrossability of bonds is therefore automatically ensured. A large body of work
has gone into deriving the effective interactions between the relatively hard beads, with iterative Boltzmann Inversion emerging as one
of the most popular techniques. When coarse-graining the interactions, technically one should include additional friction and random forces,
but fortunately in CGMD the friction is not dominant. If friction is ignored the dynamics will be too fast, usually by a factor 2 to 5.
Combined with the lower number of particles this leads to a significant computational speedup of order 100.
Agreement with detailed atomistic simulations is often recovered by a suitable time scaling. Using these techniques it is possible
to predict the rheology of melts of specific polymers up to one or two entanglement lengths.

To reach a higher degree of coarse-graining, in CGSD many more atoms are lumped together (more than
10 backbone carbon atoms), leading to relatively soft beads.
Friction and stochastic forces dominate the interactions.
Unless counteractions are taken, the soft interactions cannot prevent crossing of the bonds between the coarse-grained particles. 
Possible counteractions include application of the Twentanglement algorithm or prohibition of moves that lead to crossing of bonds.
Using such techniques it is possible to predict the rheology of melts of specific polymers up to about 10 entanglement lengths.

To predict the rheology of polymers of longer polymers (say more than 10 entanglement lengths), one is forced to make use of the
tube model by obtaining entanglement characteristics through a primitive path analysis and simulating a primitive chain network. 
Primitive path analysis algorithms allow us to determine the tube diameter and entanglement molecular weight from atomistically
detailed or CGMD simulations.
The entanglement molecular weight so obtained is often in good agreement with the entanglement molecular weight derived from
the experimental plateau modulus. 
However, predictions of time-dependent properties rely on details of the interpretation of reptation theory, and often seem to be
determined by a somewhat different entanglement molecular weight. 
A prerequisite for future computational developments is to have available a better understanding of the relation
between the two apparently different entanglement molecular weights and their relations to the effects of 
chain stiffness, finite size fluctuations and tube renewal mechanisms.
This information needs to be reliably predicted from detailed, chemically realistic models.
Improvements in understanding will be used to develop more realistic slip-link models. In this respect it is important to note
that already now slip-link models are very effective and fast in deriving the rheological properties of polymer melts.

Finally we discussed the use of super-coarse-grained models, in which a full polymer is represented by a single particle, which may
be used to predict flow and rheology on industrially relevant large time and length scales. 
Because coarse-graining has been taken to its limit, conservative interactions between full polymer chains are extremely soft
and memory or entanglement effects are prominent. To efficiently introduce memory effects we have developed 
the Responsive Particle Dynamics method.

In summary, our search for the link between chemical structure and large time dynamics has led to many innovations during the last
two decades. Computer simulations will continue to smoothen the bumpy road from chemistry to rheology.

\ack
We gratefully acknowledge support by the European Union through the Network of Excellence `SoftComp', 
the Initial Training Network `Dynacop', and NMP `Modify'.

\section*{References}


\begin{thebibliography}{100}
\expandafter\ifx\csname url\endcsname\relax
  \def\url#1{{\tt #1}}\fi
\expandafter\ifx\csname urlprefix\endcsname\relax\def\urlprefix{URL }\fi
\providecommand{\eprint}[2][]{\url{#2}}

\bibitem{DoiEdwards}
Doi M and Edwards S~F {1986} {\em {The Theory of Polymer Dynamics}\/}
  (Clarendon, Oxford)

\bibitem{Likhtman2009}
Likhtman A~E {2009} {\em {J. Non-Newtonian Fluid Mech.}\/} {\bf {157}}
  {158--161}

\bibitem{Baschnagel2000}
Baschnagel J, Binder K, Doruker P, Gusev A~A, Hahn O, Kremer K, Mattice W~L,
  Muller-Plathe F, Murat M, Paul W, Santos S, Suter U~W and Tries V {2000}
  {Bridging the gap between atomistic and coarse-grained models of polymers:
  Status and perspectives} {\em {Viscoelasticity, atomistic models, statistical
  chemistry}\/} ({\em {Adv. Polym. Sci.}\/} vol {152}) pp {41--156}

\bibitem{Kremer2002}
Kremer K and Muller-Plathe F {2002} {\em {Mol. Sim.}\/} {\bf {28}} {729--750}

\bibitem{Faller2004}
Faller R {2004} {\em {Polymer}\/} {\bf {45}} {3869--3876}

\bibitem{Aleman2009}
Aleman C, Karayiannis N~C, Curco D, Foteinopoulou K and Laso M {2009} {\em {J.
  Mol. Struct. (Theochem)}\/} {\bf {898}} {62--72}

\bibitem{Muller-Plathe2002}
Muller-Plathe F {2002} {\em {Chem. Phys. Chem.}\/} {\bf {3}} {754--769}

\bibitem{Muller-Plathe2003}
Muller-Plathe F {2003} {\em {Soft Materials}\/} {\bf {1}} {1--31}

\bibitem{Paul2004}
Paul W and Smith G~D {2004} {\em {Rep. Progr. Phys.}\/} {\bf {67}} {1117--1185}

\bibitem{Padding2007}
Padding J~T and Briels W~J {2007} {Ab-initio coarse-graining of entangled
  polymer systems} {\em {Nanostructured Soft Matter: Experiment, Theory,
  Simulation and Perspectives}\/} ed Zvelindovsky A (Springer) pp {435--458}

\bibitem{Auhl2003}
Auhl R, Everaers R, Grest G~S, Kremer K and Plimpton S~J {2003} {\em {J. Chem.
  Phys.}\/} {\bf {119}} {12718--12728}

\bibitem{Mavrantzas1999}
Mavrantzas V~G, Boone T~D, Zervopoulou E and Theodorou D~N {1999} {\em
  {Macromolecules}\/} {\bf {32}} {5072--5096}

\bibitem{Harmandaris2000}
Harmandaris V~A, Mavrantzas V~G and Theodorou D~N {2000} {\em
  {Macromolecules}\/} {\bf {33}} {8062--8076}

\bibitem{Doxastakis2001a}
Doxastakis M, Mavrantzas V~G and Theodorou D~N {2001} {\em {J. Chem. Phys.}\/}
  {\bf {115}} {11339--11351}

\bibitem{Doxastakis2001b}
Doxastakis M, Mavrantzas V~G and Theodorou D~N {2001} {\em {J. Chem. Phys.}\/}
  {\bf {115}} {11352--11361}

\bibitem{Uhlherr2001}
Uhlherr A, Mavrantzas V~G, Doxastakis M and Theodorou D~N {2001} {\em
  {Macromolecules}\/} {\bf {34}} {8554--8568}

\bibitem{Uhlherr2002}
Uhlherr A, Doxastakis M, Mavrantzas V~G, Theodorou D~N, Leak S~J, Adam N~E and
  Nyberg P~E {2002} {\em {Europhys. Lett.}\/} {\bf {57}} {506--511}

\bibitem{Karayiannis2002}
Karayiannis N~C, Mavrantzas V~G and Theodorou D~N {2002} {\em {Phys. Rev.
  Lett.}\/} {\bf {88}} 105503

\bibitem{Karayiannis2006}
Karayiannis N~C and Mavrantzas V~G {2006} {Atomistic Monte Carlo methods for
  the simulation of polymers with a linear or non-linear molecular
  architecture} {\em {Computer-Aided Chemical Engineering: Multiscale modelling
  of polymer properties}\/} ({\em {Computer-Aided Chemical Engineering}\/}
  vol~{22}) ed Laso M and Perp\`ete A (Elsevier) pp {31--67}

\bibitem{Baig2007}
Baig C and Mavrantzas V~G {2007} {\em {Phys. Rev. Lett.}\/} {\bf {99}} {257801}

\bibitem{Baig2009}
Baig C and Mavrantzas V~G {2009} {\em {Phys. Rev. B}\/} {\bf {79}} {144302}

\bibitem{Karayiannis2009}
Karayiannis N~C and Kr\"oger M {2009} {\em {Int. J. Mol. Sci.}\/} {\bf {10}}
  {5054--5089}

\bibitem{Kamio2007}
Kamio K, Moorthi K and Theodorou D~N {2007} {\em {Macromolecules}\/} {\bf {40}}
  {710--722}

\bibitem{Mulder2008a}
Mulder T, Harmandaris V~A, Lyulin A~V, van~der Vegt N~F~A, Vorselaars B and
  Michels M~A~J {2008} {\em {Macromol. Theory Sim.}\/} {\bf {17}} {290--300}

\bibitem{Mulder2008b}
Mulder T, Harmandaris V~A, Lyulin A~V, van~der Vegt N~F~A and Michels M~A~J
  {2008} {\em {Macromol. Theory Sim.}\/} {\bf {17}} {393--402}

\bibitem{Paul1995}
Paul W, Yoon D~Y and Smith G~D {1995} {\em {J. Chem. Phys.}\/} {\bf {103}}
  {1702--1709}

\bibitem{Paul1997}
Paul W, Smith G~D and Yoon D~Y {1997} {\em {Macromolecules}\/} {\bf {30}}
  {7772--7780}

\bibitem{Moore2000}
Moore J~D, Cui S~T, Cochran H~D and Cummings P~T {2000} {\em {J. Non-Newtonian
  Fluid Mech.}\/} {\bf {93}} {83}

\bibitem{Harmandaris1998}
Harmandaris V~A, Mavrantzas V~G and Theodorou D~N {1998} {\em
  {Macromolecules}\/} {\bf {31}} {7934--7943}

\bibitem{Padding2001}
Padding J~T and Briels W~J 2001 {\em J. Chem. Phys.\/} {\bf 114} 8685--8693

\bibitem{Harmandaris2003}
Harmandaris V~A, Mavrantzas V~G, Theodorou D~N, Kr\"oger M, Ramirez J, Ottinger
  H~C and Vlassopoulos D {2003} {\em {Macromolecules}\/} {\bf {36}}
  {1376--1387}

\bibitem{Smith1999}
Smith G~D, Paul W, Monkenbusch M, Willner L, Richter D, Qiu X and Ediger M~D
  {1999} {\em {Macromolecules}\/} {\bf {32}} {8857--8865}

\bibitem{Tsolou2005}
Tsolou G, Mavrantzas V~G and Theodorou D~N {2005} {\em {Macromolecules}\/} {\bf
  {38}} {1478--1492}

\bibitem{Ryckaert2005}
Ryckaert J~P {2005} {\em {Comp. Phys. Comm.}\/} {\bf {169}} {89--94}

\bibitem{Akkermans2000}
Akkermans R~L~C and Briels W~J {2000} {\em {J. Chem. Phys.}\/} {\bf {113}}
  {6409--6422}

\bibitem{Briels2002}
Briels W and Akkermans R {2002} {\em {Mol. Sim.}\/} {\bf {28}} {145--152}

\bibitem{Likos2001}
Likos C {2001} {\em {Phys. Rep.}\/} {\bf {348}} {267--439}

\bibitem{Louis2002}
Louis A {2002} {\em {J. Phys.: Condens. Matter}\/} {\bf {14}} {9187--9206}

\bibitem{Zwanzig1961}
Zwanzig R {1961} {\em Phys. Rev.\/} {\bf {124}} {983}

\bibitem{Kinjo2007}
Kinjo T and Hyodo S {2007} {\em {Phys. Rev. E}\/} {\bf {75}} {051109}

\bibitem{Harmandaris2009}
Harmandaris V~A and Kremer K {2009} {\em {Macromolecules}\/} {\bf {42}}
  {791--802}

\bibitem{Fritz2009}
Fritz D, Harmandaris V~A, Kremer K and van~der Vegt N~F~A {2009} {\em
  {Macromolecules}\/} {\bf {42}} {7579--7588}

\bibitem{Fukunaga2002}
Fukunaga H, Takimoto J and Doi M {2002} {\em {J. Chem. Phys.}\/} {\bf {116}}
  {8183--8190}

\bibitem{Harmandaris2007}
Harmandaris V~A, Reith D, Van~der Vegt N~F~A and Kremer K {2007} {\em
  {Macromol. Chem. Phys.}\/} {\bf {208}} {2109--2120}

\bibitem{Fukunaga2001}
Fukunaga H, Aoyagi T, Takimoto J and Doi M {2001} {\em {Comp. Phys. Comm.}\/}
  {\bf {142}} {224--226}

\bibitem{Tschop1998a}
Tschop W, Kremer K, Batoulis J, Burger T and Hahn O {1998} {\em {Acta
  Polymer.}\/} {\bf {49}} {61--74}

\bibitem{Reith2003}
Reith D, Putz M and Muller-Plathe F {2003} {\em {J. Comp. Chem.}\/} {\bf {24}}
  {1624--1636}

\bibitem{Abrams2003}
Abrams C~F and Kremer K {2003} {\em Macromolecules\/} {\bf {36}} {260--267}

\bibitem{Meyer2000}
Meyer H, Biermann O, Faller R, Reith D and Muller-Plathe F {2000} {\em {J.
  Chem. Phys.}\/} {\bf {113}} {6264--6275}

\bibitem{Reith2001}
Reith D, Meyer H and Muller-Plathe F {2001} {\em {Macromolecules}\/} {\bf {34}}
  {2335--2345}

\bibitem{Sun2005}
Sun Q and Faller R {2005} {\em {Comp. Chem. Engin.}\/} {\bf {29}} {2380--2385}

\bibitem{Zhao2005}
Zhao L, Li Y~G, Mi J~G and Zhong C~L {2005} {\em {J. Chem. Phys.}\/} {\bf
  {123}} {124905}

\bibitem{Faller2003}
Faller R and Reith D {2003} {\em {Macromolecules}\/} {\bf {36}} {5406--5414}

\bibitem{Sun2006a}
Sun Q and Faller R {2006} {\em {J. Chem. Theory Comp.}\/} {\bf {2}} {607--615}

\bibitem{Akkermans2001}
Akkermans R~L~C and Briels W~J {2001} {\em {J. Chem. Phys.}\/} {\bf {114}}
  {1020}

\bibitem{Ashbaugh2005}
Ashbaugh H~S, Patel H~A, Kumar S~K and Garde S {2005} {\em {J. Chem. Phys.}\/}
  {\bf {122}} {104908}

\bibitem{Nielsen2003}
Nielsen S~O, Lopez C~F, Srinivas G and Klein M~L {2003} {\em {J. Chem.
  Phys.}\/} {\bf {119}} {7043--7049}

\bibitem{Spyriouni2007}
Spyriouni T, Tzoumanekas C, Theodorou D, Mueller-Plathe F and Milano G {2007}
  {\em {Macromolecules}\/} {\bf {40}} {3876--3885}

\bibitem{Carmesin1988}
Carmesin I and Kremer K {1988} {\em {Macromolecules}\/} {\bf {21}} {2819--2823}

\bibitem{Wittmer1992}
Wittmer J, Paul W and Binder K 1992 {\em Macromolecules\/} {\bf 25} 7211--7219

\bibitem{Tanaka2000}
Tanaka M, Iwata K and Kuzuu N {2000} {\em {Comp. Theor. Polym. Sci.}\/} {\bf
  {10}} {299--308}

\bibitem{Baschnagel1991}
Baschnagel J, Binder K, Paul W, Laso M, Suter U~W, Batoulis I, Jilge W and
  B\"urger T {1991} {\em {J. Chem. Phys.}\/} {\bf {95}} {6014--6025}

\bibitem{Paul1991}
Paul W, Binder K, Kremer K and Heermann D~W {1991} {\em {Macromolecules}\/}
  {\bf {24}} {6332--6334}

\bibitem{Paul1994}
Paul W and Pistoor N 1994 {\em Macromolecules\/} {\bf 27} 1249--1255

\bibitem{Tries1997}
Tries V, Paul W, Baschnagel J and Binder K 1997 {\em J. Chem. Phys.\/} {\bf
  106} 738--748

\bibitem{Rapold1995}
Rapold R~F and Mattice W~L {1995} {\em {J. Chem. Soc., Faraday Trans.}\/} {\bf
  {91}} {2435--2441}

\bibitem{Rapold1996}
Rapold R~F and Mattice W~L {1996} {\em {Macromolecules}\/} {\bf {29}}
  {2457--2466}

\bibitem{Doruker1997}
Doruker P and Mattice W~L {1997} {\em {Macromolecules}\/} {\bf {30}}
  {5520--5526}

\bibitem{Haliloglu1998}
Haliloglu T and Mattice W~L {1998} {\em {Macromolecules}\/} {\bf {108}}
  {6898--6995}

\bibitem{Lin2006}
Lin H, Mattice W~L and Von~Meerwall E~D {2006} {\em {J. Polym. Sci. Part B:
  Polym. Phys.}\/} {\bf {44}} {2556--2571}

\bibitem{Lin2007}
Lin H, Mattice W~L and von Meerwall E~D {2007} {\em {Macromolecules}\/} {\bf
  {40}} {959--966}

\bibitem{Tschop1998b}
Tschop W, Kremer K, Hahn O, Batoulis J and Burger T {1998} {\em {Acta
  Polymer.}\/} {\bf {49}} {75--79}

\bibitem{Hess2006}
Hess B, Leon S, van~der Vegt N and Kremer K {2006} {\em {Soft Matter}\/} {\bf
  {2}} {409--414}

\bibitem{Harmandaris2006}
Harmandaris V~A, Adhikari N~P, van~der Vegt N~F~A and Kremer K {2006} {\em
  {Macromolecules}\/} {\bf {39}} {6708--6719}

\bibitem{Milano2005}
Milano G and Muller-Plathe F {2005} {\em {J. Phys. Chem. B}\/} {\bf {109}}
  {18609--18619}

\bibitem{Santangelo2007}
Santangelo G, Di~Matteo A, Muller-Plathe F and Milano G {2007} {\em {J. Phys.
  Chem. B}\/} {\bf {111}} {2765--2773}

\bibitem{Carbone2008}
Carbone P, Varzaneh H~A~K, Chen X and Mueller-Plathe F {2008} {\em {J. Chem.
  Phys.}\/} {\bf {128}} {064904}

\bibitem{Karimi-Varzaneh2008}
Karimi-Varzaneh H~A, Carbone P and Mueller-Plathe F {2008} {\em {J. Chem.
  Phys.}\/} {\bf {129}} {154904}

\bibitem{Chen2009}
Chen X, Carbone P, Santangelo G, Di~Matteo A, Milano G and Mueller-Plathe F
  {2009} {\em {Phys. Chem. Chem. Phys.}\/} {\bf {11}} {1977--1988}

\bibitem{Hahn2001}
Hahn O, Delle~Site L and Kremer K {2001} {\em {Macromol. Theory Sim.}\/} {\bf
  {10}} {288--303}

\bibitem{Faller2002}
Faller R and Muller-Plathe F {2002} {\em {Polymer}\/} {\bf {43}} {621--628}

\bibitem{Leon2005}
Leon S, van~der Vegt N, Delle~Site L and Kremer K {2005} {\em
  {Macromolecules}\/} {\bf {38}} {8078--8092}

\bibitem{Depa2005}
Depa P~K and Maranas J~K {2005} {\em {J. Chem. Phys.}\/} {\bf {123}} {094901}

\bibitem{Depa2007}
Depa P~K and Maranas J~K {2007} {\em {J. Chem. Phys.}\/} {\bf {126}} {054903}

\bibitem{Sun2006b}
Sun Q and Faller R {2006} {\em {Macromolecules}\/} {\bf {39}} {812--820}

\bibitem{Chen2006}
Chen C~X, Depa P, Sakai V~G, Maranas J~K, Lynn J~W, Peral I and Copley J~R~D
  {2006} {\em {J. Chem. Phys.}\/} {\bf {124}} {234901}

\bibitem{Chen2008}
Chen C, Depa P, Maranas J~K and Sakai V~G {2008} {\em {J. Chem. Phys.}\/} {\bf
  {128}} {124906}

\bibitem{Strauch2009}
Strauch T, Yelash L and Paul W {2009} {\em {Phys. Chem. Chem. Phys.}\/} {\bf
  {11}} {1942--1948}

\bibitem{Oettinger2007}
Oettinger H~C {2007} {\em {MRS Bulletin}\/} {\bf {32}} {936--940}

\bibitem{Qian2009}
Qian H~J, Liew C~C and Mueller-Plathe F {2009} {\em {Phys. Chem. Chem.
  Phys.}\/} {\bf {11}} {1962--1969}

\bibitem{Guenza2003}
Guenza M {2003} {\em {J. Chem. Phys.}\/} {\bf {119}} {7568--7578}

\bibitem{Guenza2008}
Guenza M~G {2008} {\em {J. Phys.: Condens. Matter}\/} {\bf {20}} {033101}

\bibitem{Hijon2010}
Hijon C, Espanol P, Vanden-Eijnden E and Delgado-Buscalioni R {2010} {\em
  {Faraday Discussions}\/} {\bf {144}} {301--322}

\bibitem{Hoogerbrugge1992}
Hoogerbrugge P~J and Koelman J~M~V~A 1992 {\em Europhys. Lett.\/} {\bf 19} 155

\bibitem{Guerrault2004}
Guerrault X, Rousseau B and Farago J {2004} {\em {J. Chem. Phys.}\/} {\bf
  {121}} {6538--6546}

\bibitem{Lahmar2007}
Lahmar F and Rousseau B {2007} {\em {Polymer}\/} {\bf {48}} {3584--3592}

\bibitem{Nikunen2007}
Nikunen P, Vattulainen I and Karttunen M {2007} {\em {Phys. Rev. E}\/} {\bf
  {75}} {036713}

\bibitem{Padding2008}
Padding J~T, Boek E~S and Briels W~J {2008} {\em {J. Chem. Phys.}\/} {\bf
  {129}} {074903}

\bibitem{vandenNoort2008}
van~den Noort A and Briels W~J {2008} {\em {J. Non-Newtonian Fluid Mech.}\/}
  {\bf {152}} {148--155}

\bibitem{Sprakel2009}
Sprakel J, Spruijt E, van~der Gucht J, Padding J~T and Briels W~J {2009} {\em
  {Soft Matter}\/} {\bf {5}} {4748--4756}

\bibitem{Kremer1988}
Kremer K, Grest G~S and Carmesin I 1988 {\em Phys. Rev. Lett.\/} {\bf 61}
  566--569

\bibitem{Kremer1990}
Kremer K and Grest G~S 1990 {\em J. Chem. Phys.\/} {\bf 92} 5057--5086

\bibitem{Gao1995}
Gao J and Weiner J~H {1995} {\em {J. Chem. Phys.}\/} {\bf {103}} {1614--1620}

\bibitem{Smith1995}
Smith S~W, Hall C~K and Freeman B~D {1995} {\em {Phys. Rev. Lett.}\/} {\bf
  {75}} {1316--1319}

\bibitem{Gao1996}
Gao J~P and Weiner J~H {1996} {\em {Macromolecules}\/} {\bf {29}} {6048--6055}

\bibitem{Smith1996}
Smith S~W, Hall C~K and Freeman B~D {1996} {\em {J. Chem. Phys.}\/} {\bf {104}}
  {5616--5637}

\bibitem{Faller1999}
Faller R, Putz M and Muller-Plathe F {1999} {\em {Int. J. Mod. Phys. C}\/} {\bf
  {10}} {355--360}

\bibitem{Aoyagi2000}
Aoyagi T and Doi M {2000} {\em {Comp. Theor. Polym. Sci.}\/} {\bf {10}}
  {317--321}

\bibitem{Putz2000}
Putz M, Kremer K and Grest G~S {2000} {\em {Europhys. Lett.}\/} {\bf {49}}
  {735--741}

\bibitem{Kroger2004}
Kr\"oger M {2004} {\em {Phys. Rep.}\/} {\bf {390}} {453--551}

\bibitem{Hou2010}
Hou J~X, Svaneborg C, Everaers R and Grest G~S {2010} {\em Phys. Rev. Lett.\/}
  {\bf {105}} {068301}

\bibitem{Padding2001a}
Padding J~T and Briels W~J 2001 {\em J. Chem. Phys.\/} {\bf 115} 2846--2859

\bibitem{Lahmar2009}
Lahmar F, Tzoumanekas C, Theodorou D~N and Rousseau B {2009} {\em
  {Macromolecules}\/} {\bf {42}} {7485--7494}

\bibitem{Padding2002}
Padding J~T and Briels W~J {2002} {\em J. Chem. Phys.\/} {\bf {117}} {925--943}

\bibitem{Padding2003}
Padding J~T and Briels W~J {2003} {\em {J. Chem. Phys.}\/} {\bf {118}}
  {10276--10286}

\bibitem{Padding2004}
Padding J~T and Briels W~J {2004} {\em {J. Chem. Phys.}\/} {\bf {120}}
  {2996--3002}

\bibitem{Kindt2005}
Kindt P and Briels W~J {2005} {\em {J. Chem. Phys.}\/} {\bf {123}} {224903}

\bibitem{Perez2010}
Perez-Aparicio R, Colmenero J, Alvarez F, Padding J~T and Briels W~J {2010}
  {\em {J. Chem. Phys.}\/} {\bf {132}} {024904}

\bibitem{Pearson1987}
Pearson D~S, Ver~Strate G, von Meerwall E and Schilling F~C {1987} {\em
  {Macromolecules}\/} {\bf {20}} {1133}

\bibitem{Faller2001}
Faller R and Muller-Plathe F {2001} {\em {Chem. Phys. Chem.}\/} {\bf {2}}
  {180--184}

\bibitem{Liu2008}
Liu H, Xue Y~H, Qian H~J, Lu Z~Y and Sun C~C {2008} {\em {J. Chem. Phys.}\/}
  {\bf {129}} {024902}

\bibitem{Spenley2000}
Spenley N~A {2000} {\em {Europhys. Lett.}\/} {\bf {49}} {534--540}

\bibitem{Kumar2001}
Kumar S and Larson R~G {2001} {\em {J. Chem. Phys.}\/} {\bf {114}} {6937--6941}

\bibitem{Hoda2010}
Hoda N and Larson R~G {2010} {\em {J. Rheol.}\/} {\bf {54}} {1061--1081}

\bibitem{Ramanathan2007}
Ramanathan S and Morse D~C {2007} {\em {J. Chem. Phys.}\/} {\bf {126}} {094906}

\bibitem{Padding2009}
Padding J~T {2009} {\em {J. Chem. Phys.}\/} {\bf {130}} {144903}

\bibitem{Everaers2004}
Everaers R, Sukumaran S~K, Grest G~S, Svaneborg C, Sivasubramanian A and Kremer
  K {2004} {\em {Science}\/} {\bf {303}} {823--826}

\bibitem{Kremer2005}
Kremer K, Sukumaran S~K, Everaers R and Grest G~S {2005} {\em {Comp. Phys.
  Comm.}\/} {\bf {169}} {75--81}

\bibitem{Kroger2005}
Kr\"oger M {2005} {\em {Comp. Phys. Comm.}\/} {\bf {168}} {209--232}

\bibitem{Tzoumanekas2006a}
Tzoumanekas C and Theodorou D~N {2006} {\em {Curr. Opin. Solid State Mat.
  Sci.}\/} {\bf {10}} {61--72}

\bibitem{Tzoumanekas2006b}
Tzoumanekas C and Theodorou D~N {2006} {\em {Macromolecules}\/} {\bf {39}}
  {4592--4604}

\bibitem{Tzoumanekas2009}
Tzoumanekas C, Lahmar F, Rousseau B and Theodorou D~N {2009} {\em
  {Macromolecules}\/} {\bf {42}} {7474--7484}

\bibitem{Hoy2009}
Hoy R~S, Foteinopoulou K and Kr\"oger M {2009} {\em Phys. Rev. E\/} {\bf {80}}
  {031803}

\bibitem{Baig2010}
Baig C and Mavrantzas V~G {2010} {\em {Soft Matter}\/} {\bf {6}} {4603--4612}

\bibitem{Sukumaran2005}
Sukumaran S~K, Grest G~S, Kremer K and Everaers R {2005} {\em {J. Polym. Sci.
  Part B: Polym. Phys.}\/} {\bf {43}} {917--933}

\bibitem{Uchidaa2008}
Uchidaa N, Grest G~S and Everaers R {2008} {\em {J. Chem. Phys.}\/} {\bf {128}}
  {044902}

\bibitem{Greco2008}
Greco F {2008} {\em {Euro. Phys. J. E}\/} {\bf {25}} {175--180}

\bibitem{Foteinopoulou2006}
Foteinopoulou K, Karayiannis N~C, Mavrantzas V~G and Kr\"oger M {2006} {\em
  {Macromolecules}\/} {\bf {39}} {4207--4216}

\bibitem{Foteinopoulou2009}
Foteinopoulou K, Karayiannis N~C, Laso M and Kr\"oger M {2009} {\em {J. Phys.
  Chem. B}\/} {\bf {113}} {442--455}

\bibitem{Schieber2003a}
Schieber J~D {2003} {\em {J. Chem. Phys.}\/} {\bf {118}} {5162}

\bibitem{Hinsch2007}
Hinsch H, Wilhelm J and Frey E 2007 {\em Euro. Phys. J. E\/} {\bf 24} 35

\bibitem{Masubuchi2001}
Masubuchi Y, Takimoto J~I, Koyama K, Ianniruberto G, Marrucci G and Greco F
  {2001} {\em {J. Chem. Phys.}\/} {\bf {115}} {4387--4394}

\bibitem{Masubuchi2003}
Masubuchi Y, Ianniruberto G, Greco F and Marrucci G {2003} {\em {J. Chem.
  Phys.}\/} {\bf {119}} {6925--6930}

\bibitem{Yaoita2004}
Yaoita T, Isaki T, Masubuchi Y, Watanabe H, Ianniruberto G, Greco F and
  Marrucci G {2004} {\em {J. Chem. Phys.}\/} {\bf {121}} {12650--12654}

\bibitem{Masubuchi2006a}
Masubuchi Y {2006} {\em {Nihon Reoroji Gakkaishi}\/} {\bf {34}} {275--282}

\bibitem{Masubuchi2006b}
Masubuchi Y, Ianniruberto G, Greco F and Marrucci G {2006} {\em {J. Non-Cryst.
  Solids}\/} {\bf {352}} {5001--5007}

\bibitem{Furuichi2007}
Furuichi K, Nonomura C, Masubuchi Y, Ianniruberto G, Greco F and Marrucci G
  {2007} {\em {Nihon Reoroji Gakkaishi}\/} {\bf {35}} {73--77}

\bibitem{Furuichi2008}
Furuichi K, Nonomura C, Masubuchi Y, Watanabe H, Ianniruberto G, Greco F and
  Marrucci G {2008} {\em {Rheol. Acta}\/} {\bf {47}} {591--599}

\bibitem{Masubuchi2009}
Masubuchi Y, Uneyama T, Watanabe H, Ianniruberto G, Greco F and Marrucci G
  {2009} {\em {Nihon Reoroji Gakkaishi}\/} {\bf {37}} {65--68}

\bibitem{Tasaki2001}
Tasaki H, Takimoto J~I and Doi M {2001} {\em {Comput. Phys. Commun.}\/} {\bf
  {142}} {136--139}

\bibitem{Doi2003}
Doi M and Takimoto J~I {2003} {\em {Philos. Trans. R. Soc. London}\/} {\bf
  {361}} {641--652}

\bibitem{Shanbhag2004}
Shanbhag S and Larson R~G {2004} {\em {Macromolecules}\/} {\bf {37}}
  {8160--8166}

\bibitem{Xu2006}
Xu F, Denn M~M and Schieber J~D {2006} {\em {J. Rheol.}\/} {\bf {50}}
  {477--494}

\bibitem{Likhtman2005}
Likhtman A~E {2005} {\em {Macromolecules}\/} {\bf {38}} {6128--6139}

\bibitem{Likhtman2007}
Likhtman A~E, Sukumaran S~K and Ramirez J {2007} {\em {Macromolecules}\/} {\bf
  {40}} {6748--6757}

\bibitem{Sukumaran2009}
Sukumaran S~K and Likhtman A~E {2009} {\em {Macromolecules}\/} {\bf {42}}
  {4300--4309}

\bibitem{Rakshit2006}
Rakshit A and Picu R~C {2006} {\em {J. Chem. Phys.}\/} {\bf {125}} {164907}

\bibitem{Rakshit2008}
Rakshit A and Picu R~C {2008} {\em {Rheol. Acta}\/} {\bf {47}} {1039--1048}

\bibitem{Murat1998}
Murat M and Kremer K {1998} {\em {J. Chem. Phys.}\/} {\bf {108}} {4340--4348}

\bibitem{McCarthy2009}
McCarty J, Lyubimov I~Y and Guenza M~G {2009} {\em {J. Phys. Chem. B}\/} {\bf
  {113}} {11876--11886}

\bibitem{Lyubimov2010}
Lyubimov I~Y, McCarty J, Clark A and Guenza M~G {2010} {\em {J. Chem. Phys.}\/}
  {\bf {132}} {224903}

\bibitem{Kindt2007}
Kindt P and Briels W~J {2007} {\em {J. Chem. Phys.}\/} {\bf {127}} {134901}

\bibitem{Briels2009}
Briels W~J {2009} {\em {Soft Matter}\/} {\bf {5}} {4401--4411}

\bibitem{Padding2010}
Padding J~T, van Ruymbeke E, Vlassopoulos D and Briels W~J {2010} {\em {Rheol.
  Acta}\/} {\bf {49}} {473--484}

\bibitem{Sprakel2011}
Sprakel J, Padding J~T and Briels W~J {2011} {\em {Europhys. Lett.}\/} {\bf
  {93}} {58003}

\bibitem{Padding2011}
Padding J~T, Mohite L~V, Auhl D, Briels W~J and Bailly C {2011} {\em {Mesoscale
  modeling of the rheology of pressure sensitive adhesives through inclusion of
  transient forces (accepted for publication in Soft Matter)}\/}

\end{thebibliography}
\providecommand{\newblock}{}

\end{document}